\documentclass[aps,prb,twocolumn,superscriptaddress,floatfix,amsmath,amssymb,10pt]{revtex4-2}
\usepackage{multirow}
\usepackage{graphicx}
\usepackage{bm}
\usepackage{dcolumn}
\usepackage{picture}
\usepackage{amsmath}
\usepackage[colorlinks=true,linkcolor=blue,urlcolor=blue,citecolor=blue]{hyperref}
\usepackage{nameref}
\usepackage{natbib}
\usepackage{times}
\usepackage{color}
\usepackage{xcolor}
\usepackage{soul}
\usepackage{amssymb}
\usepackage{booktabs}

\begin{document}

\newcommand{\STO}{SmTa\texorpdfstring{\textsubscript{7}}{7}O\texorpdfstring{\textsubscript{19}}{19}}

\title{Quantum spin liquid ground state in a rare-earth triangular antiferromagnet SmTa\texorpdfstring{\textsubscript{7}}{7}O\texorpdfstring{\textsubscript{19}}{19}}

\author{Dhanpal Bairwa}
\email[]{dhanpal@iisc.ac.in}
\affiliation{Department of Physics, Indian Institute of Science, Bangalore, 560012, India}

\author{Abhisek Bandyopadhyay}%
\email[]{abhisek.bandyopadhyay@stfc.ac.uk; abhisek.ban2011@gmail.com}
\affiliation{ISIS Neutron and Muon Source, STFC, Rutherford Appleton Laboratory, Chilton, Didcot, Oxon OX11 0QX, United Kingdom}

\author{Devashibhai Adroja}%
\email[]{devashibhai.adroja@stfc.ac.uk}
\affiliation{ISIS Neutron and Muon Source, STFC, Rutherford Appleton Laboratory, Chilton, Didcot, Oxon OX11 0QX, United Kingdom}
\affiliation{Highly Correlated Matter Research Group, Physics Department, University of Johannesburg, Auckland Park 2006, South Africa}

\author{G. B. G. Stenning }%
\affiliation{ISIS Neutron and Muon Source, STFC, Rutherford Appleton Laboratory, Chilton, Didcot, Oxon OX11 0QX, United Kingdom}

\author{Hubertus Luetkens}%
\affiliation{
PSI Center for Neutron and Muon Sciences CNM, 5232 Villigen PSI, Switzerland}

\author{Thomas James Hicken}%
\affiliation{
PSI Center for Neutron and Muon Sciences CNM, 5232 Villigen PSI, Switzerland}

\author{Jonas A. Krieger}%
\affiliation{
PSI Center for Neutron and Muon Sciences CNM, 5232 Villigen PSI, Switzerland}

\author{G. Cibin}
\affiliation{Diamond Light Source Ltd., Diamond House, Harwell Science and Innovation Campus, Didcot, Oxfordshire OX11 0DE, UK}

\author{M. Rotter}
\affiliation{McPhase Project, 01159 Dresden, Germany}
\author{S. Rayaprol}
\affiliation{UGC-DAE Consortium for Scientific Research, Mumbai Center, R-5 Shed, BARC, Trombay, Mumbai 400085, India}
\author{P. D. Babu}

\affiliation{UGC-DAE Consortium for Scientific Research, Mumbai Center, R-5 Shed, BARC, Trombay, Mumbai 400085, India}

\author{Suja Elizabeth}%
\email[]{liz@iisc.ac.in}
\affiliation{Department of Physics, Indian Institute of Science, Bangalore, 560012, India}


\begin{abstract}
The rare-earth-based geometrically frustrated triangular magnets have attracted considerable attention due to the intricate interplay between strong spin-orbit coupling and the crystal electric field (CEF), which often leads to effective spin-1/2 degrees of freedom and therefore promotes strong quantum fluctuations at low temperatures, thus offering an excellent route to stabilize a quantum spin liquid (QSL) ground state. We have investigated the ground state magnetic properties of a polycrystalline sample of $\text{SmTa}_7\text{O}_{19}$ which we propose to have a gapless QSL ground state by employing powder X-ray diffraction (XRD), X-ray absorption spectroscopy (XAS), DC and AC-magnetic susceptibility, $M$ vs. $H$ isotherm, specific heat, and muon spin rotation/relaxation measurements ($\mu$SR) down to 30 mK. The combined structural and electronic studies reveal the formation of an edge-sharing equilateral triangular lattice of Sm$^{3+}$ ions in $ab$ plane. The DC, AC magnetic susceptibility, and heat capacity measurements reveal that $\text{SmTa}_7\text{O}_{19}$ does not exhibit any long-range magnetic ordering transition down to 50 mK. The zero-field (ZF)-$\mu$SR study strongly refutes the long-range magnetically ordered ground state and/or any partial spin-freezing down to at least 30 mK. The ZF-muon-spin relaxation rate is weakly temperature dependent between 50 and 20 K, rapidly increases below $\sim$20 K and saturates at low temperatures between 2 K and 30 mK, which has been attributed to a characteristic signature of QSL systems. Further, our longitudinal-field (LF)-$\mu$SR measurements at 0.1 K reveal a dynamic nature of the magnetic ground state. In addition, our high-field specific heat data suggest a gapless nature of spin excitations in this compound.

\end{abstract}

\maketitle


\section{Introduction}

Incompatible local spin-spin exchange interactions in geometrically frustrated magnetic lattices disfavour the conventional magnetically ordered ground state even at $T$ $\rightarrow$ 0, and hence, a strongly quantum entangled fluctuating liquid-like ground state, so-called quantum spin liquid (QSL), could emerge as a result of enhanced quantum fluctuations. QSLs have been amongst the most fascinating modern day condensed matter research areas since the first theoretically introduced notion of a resonating valence bond solid in spin-1/2 triangular Heisenberg antiferromagnet by P. Anderson in 1973~\cite{Anderson-2, Anderson-1}. A QSL is a unique state of matter that exhibits characteristics such as quantum fluctuations, long-range quantum entanglement, fractionalized low-energy quasiparticle excitations (e.g. Majorana Fermions, charge neutral spinons, emergent Gauge flux, etc.), absence of long-range ordering, and also missing of spontaneous symmetry breaking \cite{wen2019experimental,savary2017quantum, Quantumspinliquids, niggemann2019classical, balents2010spin,rochner2016spin}. Materials accommodating the QSL ground state are therefore of tremendous importance from the viewpoint of both fundamental science (e.g. understanding the mechanism of high-temperature superconductivity~\cite{Anderson, Baskaran_1987}) and potential technological applications in the future generation topological quantum computation \cite{savary2017quantum,balents2010spin}. Despite a considerable research thrust in the quantum magnetism community on searching the novel QSL state of matter within the geometrically frustrated candidate materials, a true experimental realization of a QSL  phase is quite challenging, as unavoidable defects, site-disorder, and extra terms in the spin Hamiltonian often pose a strong constraint in the real materials to host the elusive QSL state experimentally ~\cite{PhysRevLett.118.087203, PhysRevLett.104.237203, PhysRevB.104.224433}.

The geometrically frustrated magnetic lattices, such as triangular, kagome, honeycomb, and hyperkagome, are prone to competing magnetic exchange interactions and hence, offer a viable ground for investigating a plethora of interesting quantum magnetic states ~\cite{moessner2006geometrical, ramirez1994strongly, MENDELS2016455, zhou2017quantum}. In particular, the triangular lattice antiferromagnets render an exemplary two-dimensional (2D) model that can maximize frustration driven quantum fluctuations and accommodate a diverse spectrum of quantum and topological phenomena, including QSL ~\cite{savary2017quantum,Quantumspinliquids, Simeng_Science_2011, Khuntia_2020,Takagi_nature_2019}. QSL behavior has been predominantly reported on the geometrically frustrated transition metal oxide systems ~\cite{PhysRevLett.106.147204, PhysRevB.103.174423, PhysRevMaterials.3.014412, Abhisek_PRB_2024, Abhisek_PRM_2024}. Nearest-neighbor antiferromagnetic interactions among the spins of a triangular lattice always impart a strong geometric constraint to prevent the lattice from simultaneously fulfilling the minimum energy condition for all magnetic bonds, thereby reducing the possibility of achieving a magnetically ordered state. When the ordering temperature is suppressed even down to 0 K due to fluctuations, a spin liquid state could emerge. The tendency towards long-range ordering can be further reduced in quantum systems containing magnetic ions with a smaller value of spin angular momentum quantum number (i.e., \(S = \frac{1}{2}\)). Spin-\(\frac{1}{2}\) compounds have smaller J$_{eff}$ state, and therefore, according to Heisenberg's uncertainty principle, exhibit smaller precision in projection measurements, leading to enhanced fluctuations and contributing to the quantum mechanical ground states. Conversely, for the compounds with spin greater than \(\frac{1}{2}\), quantum fluctuations get reduced, exposing the system to stronger magnetic interactions and causing an increased probability of displaying magnetic ordering. 

Apart from the transition metal oxide-based systems, several rare-earth-based geometrically frustrated magnetic lattices have proven to be fertile ground for exploring QSL candidacy. A few examples include NdTa$_7$O$_{19}$ \cite{arh2022ising}, NaYbO$_2$ \cite{bordelon2019field}, NaYbSe$_2$ \cite{ranjith2019anisotropic}, Ce$_2$Zr$_2$O$_7$ \cite{gao2019experimental}, Tb$_2$Ti$_2$O$_7$ \cite{takatsu2011quantum}, and Li$_3$Yb$_3$Te$_2$O$_{12}$ \cite{PhysRevB.106.104404}. In all these systems, the magnetic rare-earth ions exhibit spins larger than \(\frac{1}{2}\). While rare-earth ions possess relatively large total angular momentum \(J\), the degeneracy (\(2J + 1\)) is typically lifted by the crystal electric field (CEF). As a result, the lowest energy state often forms a Kramers doublet for an odd number of 4\textit{f}-electrons (in the paramagnetic state), with a substantial energy gap separating it from the first excited state. Consequently, the system behaves effectively as spin-\(\frac{1}{2}\), which can be subjected to enhanced quantum fluctuations and consequently, a QSL state \cite{li2016anisotropic}. Additionally, non-Kramer's ions (even number of f electrons) are also shown to display \(S=\frac{1}{2}\) characteristics at low temperatures \cite{clark2019two, mukherjee2014effective, bu2022gapless}. Consequently, the rare-earth-based frustrated magnets have been proposed as suitable candidates for exploring the quantum materials candidacy and exotic quantum magnetic ground states. 

$RE$Ta$_7$O$_{19}$ ($RE$ = rare-earth) is an interesting new addition in the family of rare-earth based frustrated triangular magnets where the trivalent magnetic `$RE$ cations form an edge-sharing triangular network in the $ab$-plane with the nonmagnetic Ta$^{5+}$ (5$d^0$) ions being positioned in between, thus fulfilling the criteria for geometric frustration \cite{WANG2023168390, rossell1976unit, zuev1991xray}. Here, we present a comprehensive experimental study on the magnetic ground state of the Sm$^{3+}$ member of this family, SmTa$_7$O$_{19}$, using in-depth DC, AC magnetic susceptibility, $M-H$ isotherm, specific heat, thermodynamic scaling analysis, and zero-field (ZF) and longitudinal-field (LF) $\mu$SR characterizations. It is important to point out in this context that Sm$^{3+}$ ion (4$f^5$: $L$ = 5, $S$ = 5/2) has a spin-orbit-split total angular momentum $J$ = 5/2 ground state, and hence, (2$J$+1) = 6-fold degeneracy. Now considering odd number (5) of 4$f$ electrons, the Sm$^{3+}$ will have three Kramers doublet under crystal-electric-field potential with a point symmetry lower than cubic. The Sm-$L_2$ and $L_3$ edge X-ray absorption near edge structure (XANES) spectroscopies infer the stoichiometrically desired 3+ valence of Sm in this compound. Despite having finite antiferromagnetic correlation ($\Theta_W \sim$ -0.4 to -0.7 K estimated from the low-$T$ Curie-Weiss fits) among the Sm$^{3+}$ moments, our combined bulk DC and AC magnetic susceptibility, specific heat, and $\mu$SR characterizations demonstrate the absence of long-range magnetic ordering and/or a frozen magnetic ground state in SmTa$_7$O$_{19}$ down to the lowest measured 0.03 K. Rather, the system manifests persistent spin dynamics, as evidenced from our LF-$\mu$SR data analysis at 0.1 K. This gives rise to a high frustration index, $f = |\frac{\Theta_W}{T_{min}}|$ (where $T_{min}$ is the lowest temperature down to which no magnetic transition is observed) $>$ 13, which originates from the edge-sharing equilateral triangular network of Sm$^{3+}$, imparting enhanced quantum fluctuations among the Sm-moments, evading the system from a magnetic phase transition, and hence, stabilizing a quantum entangled dynamic QSL ground state in SmTa$_7$O$_{19}$. In addition, failure of universal scaling relation and data collapse of the $\chi_{DC}(T)$, $M(H)$ and $C_m/T$ in $T/H$, $H/T$ and $T/H$, respectively, refute a random-singlet state (RSS) in our SmTa$_7$O$_{19}$, likely in agreement with the absence of quenched disorder (by means of site-disorder or charge-disproportionation of the magnetic ion) in our system. Finally, the power-law behavior of the high-field $C_m(T)$ data at low temperatures suggests the gapless nature of spin excitations in SmTa$_7$O$_{19}$.

\section{Experimental techniques}
$\text{SmTa}_7\text{O}_{19}$ was synthesized by solid state reaction method. $\text{Sm}_2\text{O}_3$ and $\text{Ta}_2\text{O}_5$ were preheated at 800 °C to remove moisture. The precursors were then weighed in stoichiometric ratio and thoroughly ground to ensure uniform mixing. Subsequently, the resultant powder was compressed into pellets and sintered at 1400°C. This sintering procedure was repeated five times with intermittent grinding and pelletization. The phase purity was evaluated using powder X-ray diffraction (XRD) measurement conducted with a Rigaku instrument equipped with \textit{Cu-K$\alpha$} radiation ($\lambda = 1.5406$ Å) at room temperature. The refined crystal structure was obtained through a structural refinement performed by the Rietveld technique using FULLPROF ~\cite{rietveld1969, rodriguez-carvajal}. Sm-$L_2$ and $L_3$ edge X-ray absorption spectroscopy (XAS) measurements were carried out at the B-18 beamline of the Diamond Light Source, UK, in
standard transmission geometry at room temperature. 
DC magnetic susceptibility and isothermal magnetization measurements were conducted using a Quantum Design MPMS3 superconducting quantum interference device (SQUID) equipped with
a vibrating sample magnetometer in the temperature range of
2-300 K and in applied magnetic fields $H$ of up to $\pm$ 60 kOe. Specific heat measurements in the 2-300 K $T$-range were performed using a physical property measurement system (PPMS, Quantum Design). In addition, specific heat and AC magnetic susceptibility data were also collected between 0.05 and 4 K using a Quantum Design Dynacool PPMS at the Materials Characterization Laboratory (MCL) of ISIS Facility, UK. Muon spin rotation/relaxation experiments in both the zero-field and longitudinal-field modes were performed in the FLAME spectrometer at the Swiss muon source of PSI, Switzerland. The measurements were performed on an annealed pressed pellet of 13 mm in diameter and about 1.5 mm thick, which was loaded in the Variox plus Kevinox dilution fridge insert using a Cu-holder to achieve the temperature range between 0.03 and 50 K. The collected $\mu$SR data were fitted using the musrfit program~\cite{Musrfit_2012}.

\section{Results and Discussions}
\subsection{Structural characterization}
\begin{figure}
  \includegraphics[width= \columnwidth, trim={100 500 100 100},clip]{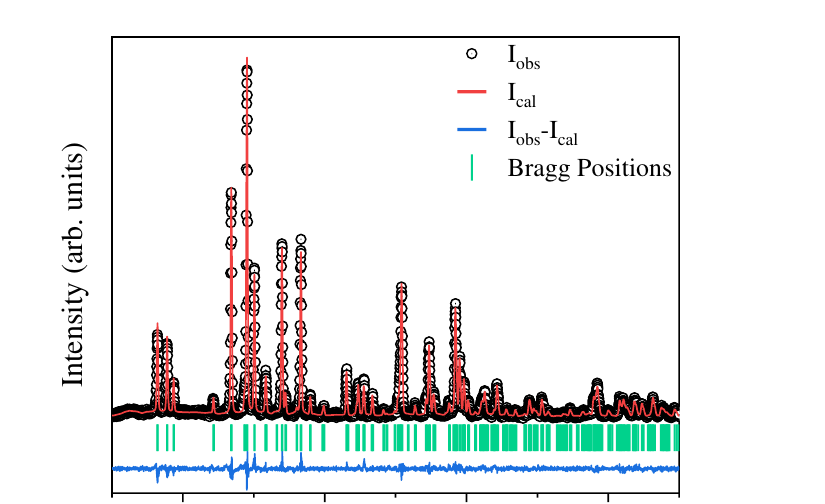} 
  \caption{Rietveld refined powder X-ray diffraction data of $\text{SmTa}_7\text{O}_{19}$ at room temperature with experimental and calculated data shown by black open circle and red line, respectively. The Bragg positions and difference between the experimental and calculated data are represented by green vertical bars and blue line, respectively.}
  \label{STO PXRD}
\end{figure}

\begin{figure}

  \includegraphics[width= \columnwidth, trim={0 0 00 0},clip]{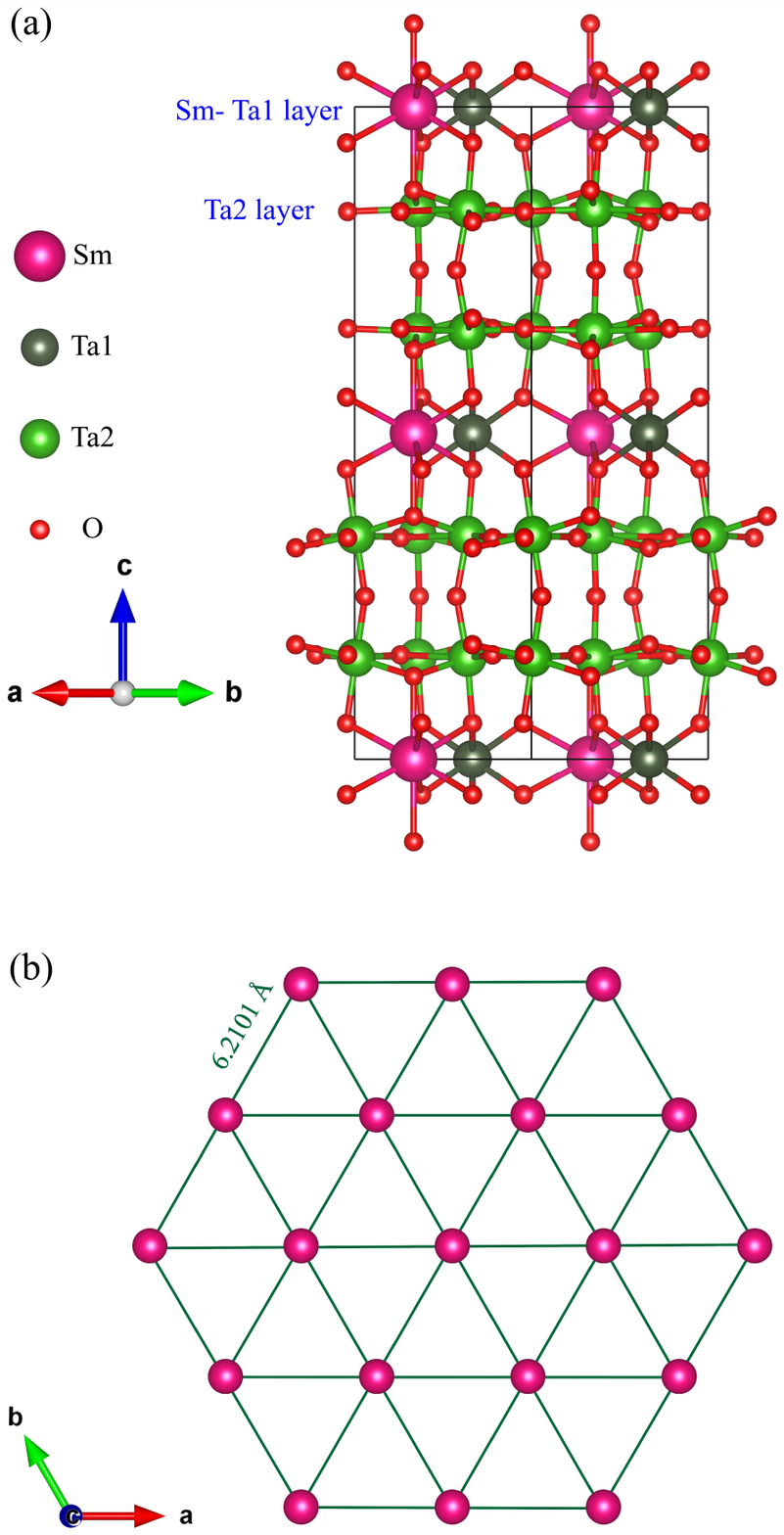} 
  \caption{ (a) Crystal structure of $\text{SmTa}_7\text{O}_{19}$ in  \( P\bar{6}c2 \) space group. (b) Triangular lattice formed by Sm$^{3+}$ ions in the ab plane. These Sm layers are separated by a distance of c/2 along the c-axis.}
  \label{STO structure}
\end{figure}

The Rietveld refinement of the room temperature powder XRD data from SmTa$_7$O$_{19}$ is shown in Fig. \ref{STO PXRD}, which clearly reveals the pure single phase nature of the sample with the hexagonal \( P\bar{6}c2 \) (\(   No.~188 \)) space group. The refined structural parameters along with the goodness-of-fit factors are summarized in Table~\ref{STO:atomic_positions}. Fig. \ref{STO structure} illustrates the refined crystal structure of $\text{SmTa}_7\text{O}_{19}$ \cite{momma2011vesta}. The unit cell contains two formula units. It adopts a layered structure, where the Sm1-Ta1 layer is sandwiched between two Ta2 layers. The distance between the Sm1-Ta1 layers along the c direction is \textit{c}/2, a relatively large separation. Within the Sm1-Ta1 layer, the Sm1 and Ta1 ions maintain an ordered arrangement, as depicted in Fig. \ref{STO structure} (a). The distance of separation between the Sm ions in the ab plane is equal to lattice parameter $a$ (= $b$). The substantially large distance between Sm layers along the c direction classifies it as a pseudo-2D material, where interactions along the c direction are relatively small. The Sm atom resides at the center of the \( \text{SmO}_8 \) dodecahedron, which is a distorted cubic structure with 12 faces. In the Sm1-Ta1 layer, the Ta1 atoms occupy the center of the \( \text{TaO}_6 \) octahedra, while in the Ta2 layer, the atoms ( Ta2 ) are at the center of the \( \text{TaO}_7 \) pentagonal bipyramid. Furthermore, the Sm ions of this structure constitute an edge-sharing highly geometrically frustrated equilateral triangular network in the $ab$-plane [see Fig.~\ref{STO structure} (b)].

\begin{table}[tbh!]
\centering
\caption{The structural parameters of $\text{SmTa}_7\text{O}_{19}$ were determined from Rietveld refinement of room-temperature powder X-ray diffraction data. The crystal structure, characterized by the \( P\bar{6}c2 \) (\( 188 \)) space group, has lattice parameters: $a = b = 6.21015(12)$ Å, $c = 19.8880(5)$ Å, \( \alpha = \beta = 90^\circ \), \( \gamma = 120^\circ \), and \( V = 664.24(2) \) Å\(^3\). The refinement parameters are as follows: \( R_p = 15\% \), \( R_{wp} = 12.7\% \), \( R_{exp} = 9.19\% \), and \( \chi^2 = 1.89 \). }

\begin{tabular}{c c c c c c c}
\hline
Atom & Site & $x$ & $y$ & $z$ & $B_{\text{iso}}$ & Occ\\
\hline
Sm1 & 2c & $1/3$ & $2/3$ & $0.0$ & 0.1134  & $1$ \\
Ta1 & 2e & $2/3$ & $1/3$ & $0.0$ & 0.1348 & $1$ \\
Ta2 & 12i & $0.3545(6)$ & $0.3620(4)$ & $0.1559(1)$ & 0.1227 & $1$ \\
O1  & 12i & $0.259(3)$ & $0.04(5)$ & $0.1603(7)$ & 0.2598 & $1$ \\
O2  & 12i & $0.378(3)$ & $0.045(4)$ & $0.9448(5)$ & 0.2895 & $1$ \\
O3  & 6k  & $0.365(5)$ & $0.424(4)$ & $1/4$ & 0.2277 & $1$ \\
O4  & 4i  & $2/3$ & $1/3$ & $0.1758(13)$ & 0.2954 & $1$ \\
O5  & 4h  & $1/3$ & $2/3$ & $0.1271(17)$ & 0.2388 & $1$ \\
\hline
\end{tabular}
\label{STO:atomic_positions}
\end{table}

\subsection{Electronic characterization}
\begin{figure}[tbh!]
{\includegraphics[width= \columnwidth, trim={50 320 50 70},clip]{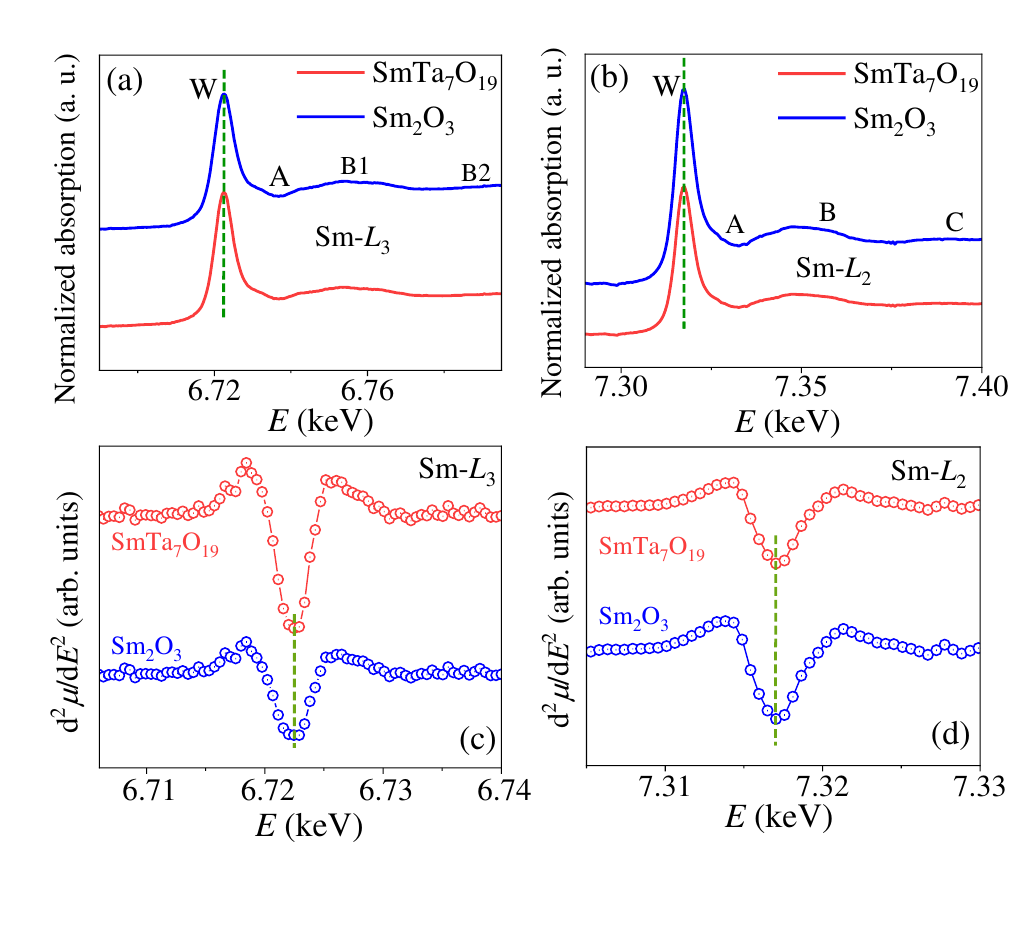}}
\caption{ (Color online) Sm-$L_3$ (a) and $L_2$ (b) edge XANES spectra of SmTa$_7$O$_{19}$ and Sm$_2$O$_3$. (c) and (d) represent the respective second derivative curves.}
\label{FIG:XANES}
\end{figure}

The Sm-valence in SmTa$_7$O$_{19}$ was checked using X-ray absorption near edge structure (XANES) spectroscopy at the Sm-$L_3$ and $L_2$ edges. The results are summarized in Figs. \ref{FIG:XANES} (a) - (d). In addition, we have also measured the Sm-$L_3$ and $L_2$ edge XANES on the Sm$_2$O$_3$ reference for comparison. As displayed in Figs. \ref{FIG:XANES} (a) and (b), both the $L_3$ and $L_2$-XANES spectra show an intense peak, characteristic of the white line feature (W) of Sm$^{3+}$ due to 2$p_{3/2}$ to 5$d_{3/2, 5/2}$ and 2$p_{1/2}$ to 5$d_{3/2}$ transitions, respectively \cite{SmLxanesprb2012}. The other near-edge features are denoted by A, B1, B2, and A, B, C, respectively, above the white line of $L_3$ and $L_2$ edges. The observed spectral characteristics, including peak shape, asymmetry, and energy positions of the white line as well as other weak near-edge features of both the $L_3$ and $L_2$-edges XANES spectra of SmTa$_7$O$_{19}$ match perfectly with those of the Sm$_2$O$_3$ reference. This is likely in agreement with the stoichiometrically expected 3+ valence of Sm in SmTa$_7$O$_{19}$. Moreover, the peak shape and energy positions of the second derivative curves of the respective white line spectra of both the $L_3$ and $L_2$ edges of SmTa$_7$O$_{19}$ align well with those of the Sm$_2$O$_3$, reconfirming the 3+ charge state of Sm in our SmTa$_7$O$_{19}$ system. 

Finally, in absence of a suitable Sm$^{2+}$ reference for the XANES measurement, we have digitized the Sm-$L_3$ edge XANES spectra of SmCl$_2$ (Sm$^{2+}$), SmF$_3$(Sm$^{3+}$) and Sm$_3$S$_4$ (mixed Sm$^{2+}$/Sm$^{3+}$) \cite{Hu-JALCOM-1997}, and plotted together with our experimentally collected Sm-$L_3$ edge XANES data in the same plot [see Fig.~\ref{FIG:AP1} in the Appendix~\ref{XANES}] for comparison. As shown, the white-line of the Sm-$L_3$ XANES of SmTa$_7$O$_{19}$ exactly matches with that of SmF$_3$ (Sm$^{3+}$), while shows no feature at/near the peak position of the SmCl$_2$ (Sm$^{2+}$) XANES spectrum. This clearly refutes 2+ valence of Sm and supports further the 3+ oxidation state of Sm in our compound. 

\subsection{DC and AC magnetizations and power-law scaling}

\begin{figure*}[tbh!]
{\includegraphics[width=1.00\linewidth, trim={50 80 50 00},clip]{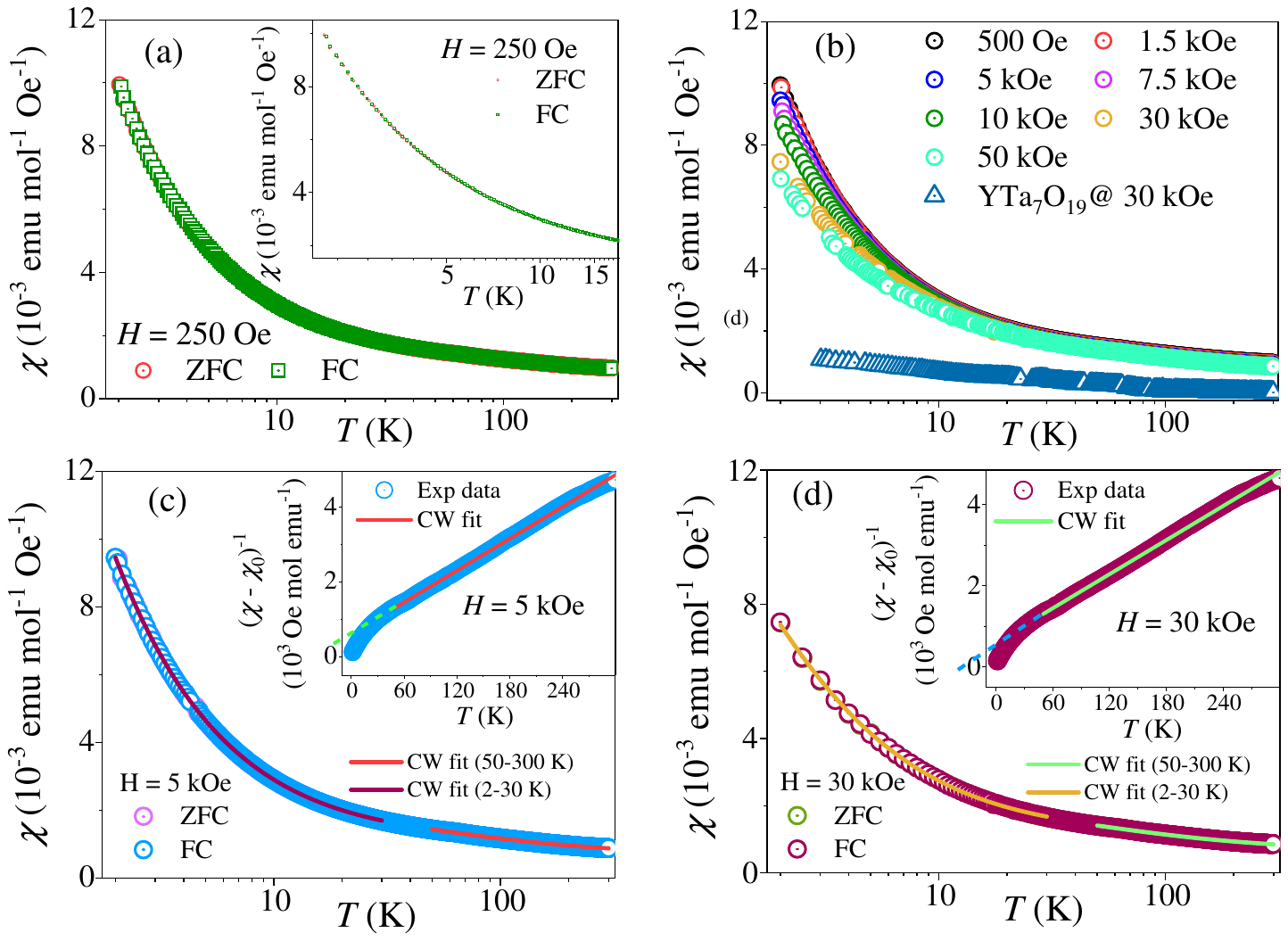}}
\caption{(Color online) (a) zero-field-cooled (ZFC) and field-cooled (FC) DC magnetic susceptibility curves as a function of temperature in 250 Oe field; Inset: Zoom-in view of the same. (b) $T$-dependence of the field-cooled DC magnetic susceptibility curves in various applied magnetic fields; further, 30 kOe field-cooled DC magnetic susceptibility curve of YTa$_7$O$_{19}$ is also plotted in the same figure as a nonmagnetic reference for comparison. Temperature dependence of ZFC and FC DC magnetic susceptibilities in (c) 5 kOe and (d) 30 kOe applied fields, along with the respective Curie-Weiss (CW) fits in both high-$T$ (50-300 K) and low-$T$ (2-30 K) regions; Insets: Corresponding inverse susceptibility versus temperature plots along with the linear CW fits for 50-300 K $T$-range.}
\label{FIG:DC_CHI}
\end{figure*}

\begin{figure*}
{\includegraphics[width=1.00\linewidth, trim={00 450 20 70},clip]{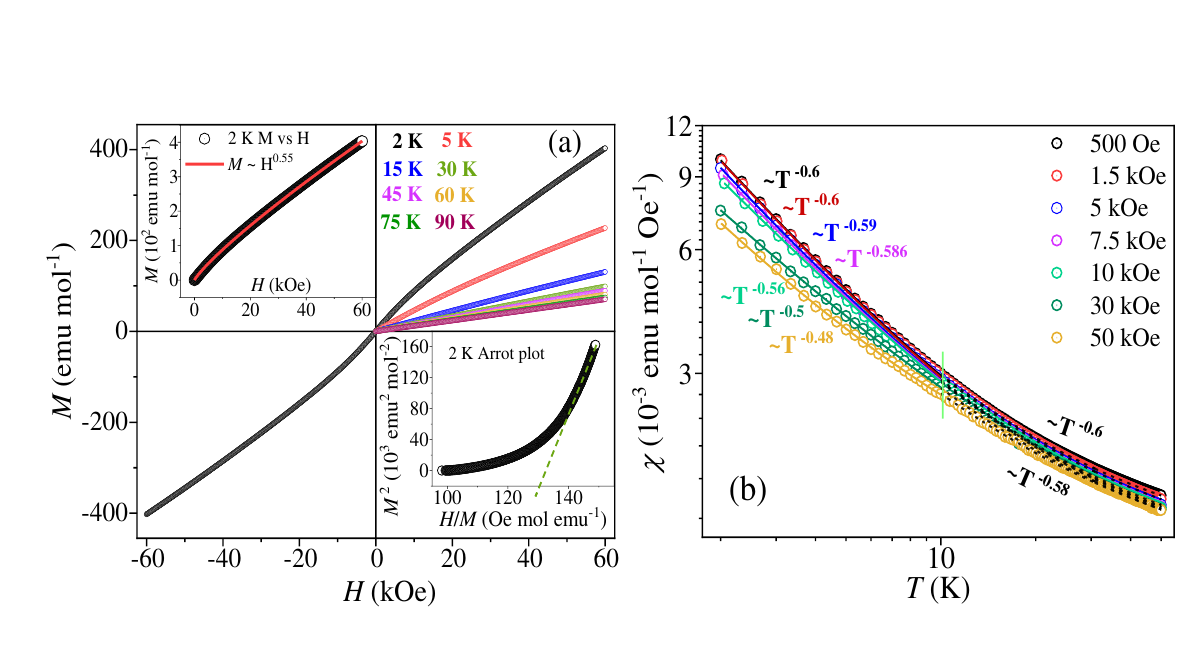}}
\caption{(Color online) (a) Field dependent isothermal magnetization data at some selected temperatures; Bottom right inset: 2 K $M^2$ versus $H/M$ Arrot plot; Top left inset: 2 K $M$-$H$ isotherm (shown only for the first quadrant) along with a power-law fit, $M(H) \propto H^{1-\alpha_m}$ with $\alpha_m$ $\approx$ 0.45. (b) $T$-dependence of the field-cooled DC magnetic susceptibility curves in different applied magnetic fields on a log-log scale, along with power-law fits (solid colored lines for 2 - 10 K and dashed black lines for 10 - 50 K).}
\label{FIG:Scaling}
\end{figure*}

As depicted in Figs.~\ref{FIG:DC_CHI} (a), (c) - (d), and Appendix~\ref{Magnetic data}, the zero-field-cooled (ZFC) and field-cooled (FC) DC susceptibility data in different applied magnetic fields, ranging from 250 Oe to 50 kOe, reveal featureless paramagnetic-like susceptibility behavior without any ZFC/FC bifurcation, possibly suggesting absence of a long-range magnetic ordering transition or a spin freezing transition in this system down to 2 K. The $M$ - $H$ isotherms, collected at several different temperatures as displayed in Fig.~\ref{FIG:Scaling} (a), clearly refute any signature of hysteresis through the absence of coercivity and remanent magnetization down to 2 K. This, along with the negative intercept of the $M^2$ versus $H/M$ Arrot plot on the $M^2$-axis [see bottom right inset of Fig.~\ref{FIG:Scaling} (a)] at 2 K , further supports that our SmTa$_7$O$_{19}$ material does not possess any spontaneous ferromagnetic component.

\begin{figure}[tbh!]
{\includegraphics[width=1.00\linewidth,  trim={100 450 100 70},clip]{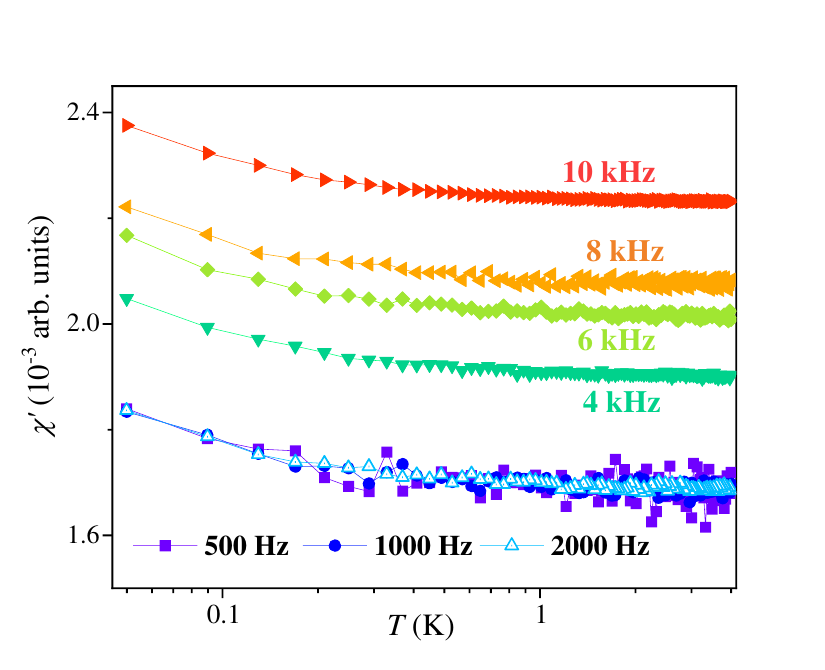}}
\caption{(Color online) Temperature dependence of real part of the ac magnetic susceptibility at some selected frequencies in zero DC bias and 1 Oe excitation field.}
\label{FIG:AC_Chi}
\end{figure}

Next, the temperature dependence of AC magnetic susceptibility of SmTa$_7$O$_{19}$ was measured at various frequencies down to 0.05 K. As shown in Fig.~\ref{FIG:AC_Chi}, the real part of the AC magnetic susceptibility, $\chi^{\prime}$($T$), exhibits neither a cusp/peak-like feature nor any noticeable frequency- dependence anomaly between 4 K and 0.05 K, thus ruling out a long-range magnetically ordered state and/or a spin-frozen magnetic ground state down to 0.05 K in SmTa$_7$O$_{19}$.   

At this point, it is important to stress that in such disordered apparently paramagnetic systems, the applicability of the Curie-Weiss (CW) law and the resulting CW fit parameters often carry some uncertainties based on the choice of the applied magnetic field and the temperature range of CW fitting. In our SmTa$_7$O$_{19}$, the feasibility of CW fitting has been validated by plotting the measured field-independent high-field ($>$ 5 kOe) FC DC susceptibility data over the chosen temperature region of fitting, as illustrated in Fig.~\ref{FIG:DC_CHI} (b). The 30 kOe FC DC $\chi$ data of YTa$_7$O$_{19}$ was also measured and shown in the same Fig.~\ref{FIG:DC_CHI} (b) just as a nonmagnetic reference of our SmTa$_7$O$_{19}$. Accordingly, the CW fits [$\chi = \chi_0 + \frac{C}{(T - \Theta_W)}$, where $\chi_0$ being a temperature-independent susceptibility, and $C$ and $\Theta_W$ are the Curie constant and Weiss temperature, respectively] were performed on the 5, 7.5, 10, 30, and 50 kOe FC DC susceptibility data of SmTa$_7$O$_{19}$ between 50 and 300 K, the results of which are shown in Figs.~\ref{FIG:DC_CHI} (c), (d) and Appendix~\ref{Magnetic data}. These fits yield an effective magnetic moment, $\mu_{eff}$ = 0.75 to 0.76 $\mu_B$/Sm and Weiss temperature, $\Theta_W$, varying only between -42 and -44 K depending on the choice of applied fields. The estimated $\mu_{eff}$ of $\sim$ 0.75-0.76 $\mu_B$ per Sm ion is much closer to the free Sm$^{3+}$ ion (4$f^5$, ${}^{6}$$H_{5/2}$; ($\mu^{free}_{eff}$)$_{Sm^{3+}}$ = 0.85 $\mu_B$). The obtained large negative value of $\Theta^{HT}_W$ ($\sim$ -42 to -44 K) possibly suggests the antiferromagnetic exchange energy scale. The slight mismatch between the inverse susceptibility data and the linear CW fit at temperatures above $\sim$ 250 K could be due to the small energy separation between the $J_{eff}$ = 5/2 ground and $J_{eff}$ = 7/2 excited states of a free Sm$^{3+}$ ion. As evident in Figs.~\ref{FIG:DC_CHI} (c), (d), and Appendix~\ref{Magnetic data}, there is a clear slope-change in the inverse DC magnetic susceptibility data below $\sim$ 50 K, which is due to the change of the population of the Sm-CEF levels upon temperature lowering. The CW law was also applied to fit the low-temperature (2-30 K) FC DC magnetic susceptibility data and the fitting results are demonstrated in Figs.~\ref{FIG:DC_CHI} (c), (d), and Appendix~\ref{Magnetic data} for different applied fields. Such a low-$T$ CW fit gives rise to an effective magnetic moment, $\mu_{eff}^{LT}$ $\sim$ 0.38 $\mu_B$/Sm, which is much smaller than the theoretically predicted moment for a free Sm$^{3+}$ ion, possibly suggesting emergence of a crystal-electric-field-split low-energy $J_{eff}$ state 
governing the low-$T$ physics. The corresponding Weiss temperature, $\Theta_W^{LT}$, varying between -0.4 and -0.7 K, refers to the weak antiferromagnetic exchange between the Sm$^{3+}$ moments at low temperature. So, it is quite clear that the crystal electric field has a strong impact on the underlying magnetic ground state properties of SmTa$_7$O$_{19}$. Our point charge model CEF calculation indicated a CEF doublet ground state, first excited CEF doublet at 24 meV and second excited CEF doublet at 39 meV.  Considering the fact that Sm has very high neutron absorption cross section (5922 barn) and $^{154}\text{Sm}$ isotope is highly expensive, we refrain from conducting a reliable inelastic neutron scattering study in this work for directly confirming the Sm-CEF-level scheme predicted from the point charge model. However, we have also fitted the 10 kOe field-cooled dc susceptibility data using the CEF model [$\chi_{CEF} = \frac{C}{T-\Theta_W}\times\frac{5+26\exp(\frac{-\Delta}{k_BT})+32[1-\exp(\frac{-\Delta}{k_BT})](\frac{k_BT}{\Delta})}{21[1+2\exp(\frac{-\Delta}{k_BT})]}$] \cite{CEFfit_Susc} over the full $T$ range. As displayed in Fig.~\ref{FIG:AP3} in the Appendix~\ref{Magnetic data}, Our fit result reveals an energy gap, $\Delta$, between the ground and first excited state Kramers doublet to be about 287 K ($\approx$ 24.7 meV). Clearly, this is in agreement with our point charge model calculation. The estimated small negative $\Theta_W^{CEF}$ ($\sim$ -0.5 K) therefore suggests weak antiferromagnetic exchange between the CEF-split low-energy $J_{eff}$ states of Sm$^{3+}$ in SmTa$_7$O$_{19}$.

The observed variation of the Weiss temperature, obtained from the CW fittings, as a function of the applied magnetic field in absence of any spin ordering/freezing is quite common in the reported other QSL materials~\cite{Kumar_PRB_2016}.
As our combined DC and AC magnetization measurements reveal no magnetic transition or anomaly down to at least 0.05 K, the frustration index, $f = |\frac{\Theta_W}{T_N}|$ where $T_N$ being the lowest measured temperature = 0.05 K, is greater than 13, reflecting highly frustrated geometry of the magnetic lattice, which consists of the edge-sharing equilateral triangles of Sm$^{3+}$. It is to be pointed out that the relative higher Sm-Sm distance of SmTa$_7$O$_{19}$, compared to the rare-earth based several other triangular lattice QSL candidates [see Table~\ref{tab:QSL materials}], possibly gives rise to weaker magnetic exchanges between the Sm moments, resulting in the relative weakening of the geometric frustration in our SmTa$_7$O$_{19}$ with respect to the triangular lattice based rare-earth QSL materials as clearly demonstrated in Table~\ref{tab:QSL materials}.  

Now, in order to gain further insights, we closely inspect the low-temperature $\chi_{DC}$ data. Fig.~\ref{FIG:Scaling} (b) shows the temperature dependence of DC magnetic susceptibility in applied several magnetic fields in the $T$-range of 2-50 K on a log-log scale. Above 50 K, $\chi(T)$ is nearly field-independent, while below 50 K, $\chi(T)$ is gradually suppressed with increasing field. Strikingly, two different power-law regimes above and
below 10 K have been envisaged. $\chi(T)$ follows the power-law dependence $T^{-\alpha_s}$ with $\alpha_s$ = 0.58 - 0.6 for $T >$ 10 K and it decreases to $\alpha$ = 0.48 - 0.6 for $T <$ 10 K upon varying applied field. This transition to a stronger sub-Curie behavior with temperature lowering implies finite spin degrees of freedom and the development of abundant low-energy states, reflecting the change of spin-spin correlations in SmTa$_7$O$_{19}$ \cite{Abhisek_PRB_2024,Choi_PRL_2019,Suheon_PRR_2024}. The observed sub-Curie behavior has been widely reported in the systems with random/frustrated magnetism or competing exchange interactions \cite{Abhisek_PRB_2024,Choi_PRL_2019,Suheon_PRR_2024,Castro_PRL_1998,Hirsch_PRB_1980,Sungwon_PRB_2018}. We now turn to the isothermal magnetization data displayed in Fig.~\ref{FIG:Scaling} (a). It is evident that below 15 K, $M-H$ isotherms deviate from linearity and do not follow a paramagnetic Brillouin function-like behavior. Rather at 2 K, the $M(H)$ curve maintains a power-law dependence, $H^{1-\alpha_m}$ with $\alpha_m$ = 0.45 over an entirely measuring field range [see inset to Fig.~\ref{FIG:Scaling} (a)], suggesting intrinsic spin contributions at low-$T$. Together with the sub-Curie behavior of $\chi(T)$, the sub-linear/power-law dependence of $M(H)$ with similar scaling exponents stands for a characteristic signature of a spin-liquid/valence-bond/random-singlet ground state~\cite{Abhisek_PRB_2024,Choi_PRL_2019,Suheon_PRR_2024,Lee_PRB_2023,Kimchi_Nature_2018,Kimchi_PRX_2018} and has been widely discussed in the context of a range of frustrated quantum magnets~\cite{Abhisek_PRB_2024,Choi_PRL_2019,Suheon_PRR_2024,Lee_PRB_2023,Kimchi_PRX_2018,Janod_PRB_2001,Choi_PRB_2004,DO_PRB_2014}.   

In contrary to the previous study by Wang et al \cite{WANG2023168390} that Van Vleck paramagnetism could sufficiently explain the low-$T$ magnetic susceptibility and $M-H$ isotherm, the stronger sub-Curie dependence of $\chi(T)$ at low-$T$ ($<$10 K) in conjunction with the gradual suppression of $\chi$ with increasing applied field, and the power-law behavior of 2 K $M$ vs $H$ isotherm over the measuring entire field range, clearly indicate presence of finite spin degrees of freedom and that the intrinsic spin correlations, not only the Van Vleck type, govern the low-$T$ magnetic response in our SmTa$_7$O$_{19}$. In addition, the applicability of the CEF model in correctly describing the low temperature ($<$ 15 K) bulk dc susceptibility data [see Fig.~\ref{FIG:AP3} in Appendix~\ref{Magnetic data}] further ensures that Van Vleck paramagnetism is no longer valid to explain the low-$T$ magnetic behavior.

\subsection{Muon spin rotation/relaxation}

To identify the true nature of the magnetic ground state, as well as probe the local spin dynamics, one needs to employ a local microscopic magnetic probe.  Muon spin rotation/relaxation ($\mu$SR) is an extremely sensitive microscopic local probe to detect static small local fields (down to approximately 0.1 Oe) arising from weak long-range magnetic order or spin freezing. $\mu$SR can also be used as the most powerful local magnetic technique by far to precisely determine the nature of the local spin dynamics and internal magnetic fields of any magnetically disordered material~\cite{Tushar_PRB_2017,Abhisek_PRB_2022,Kenney_PRB_2019}. We therefore preformed zero-field (ZF) and longitudinal-field (LF) muon spin rotation/relaxation ($\mu$SR) measurements on SmTa$_7$O$_{19}$.

\begin{figure*}[tbh!]
{\includegraphics[width=1.0\linewidth,  trim={00 300 20 50},clip]{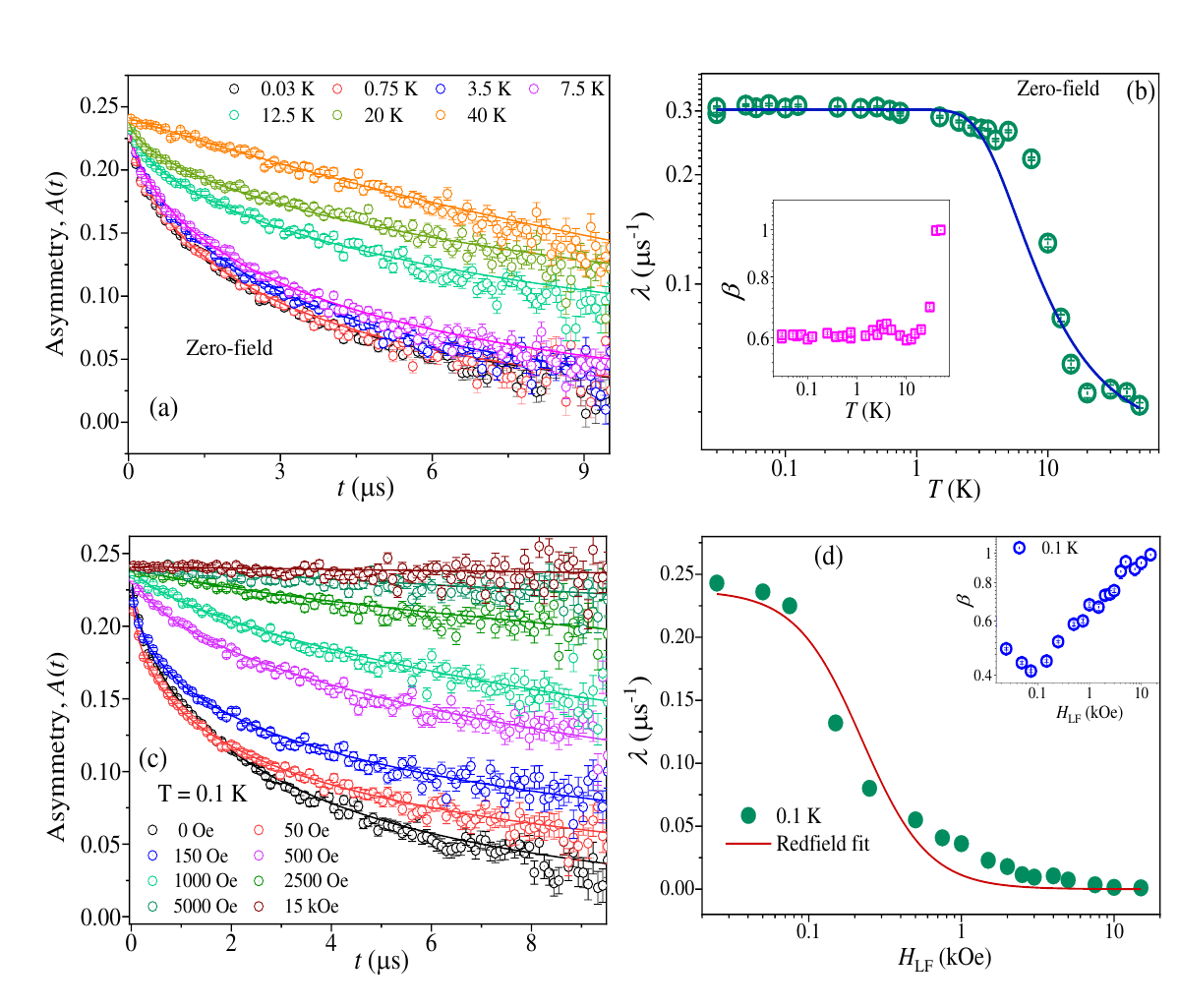}}
\caption{(Color online) (a) Time evolution of the zero-field muon asymmetry spectra (colored circles) at selected temperatures, along with fits (solid colored
lines). (b) Temperature dependence of the ZF-muon spin relaxation rate on a log-log scale along with the phenomenological Orbach model fit (blue solid line); Inset: Corresponding stretched exponent vs. temperature variation in a log-log scaled plot. (c) Time evolution of the muon asymmetry spectra at $T$ = 0.1 K in applied several longitudinal fields. (d) Longitudinal field (LF) dependence of relaxation rate, $\lambda$, along with the Redfield model fitting (solid red line) on a log-linear scale; Inset: LF dependence of stretched exponent in a log-log scale.}
\label{FIG:MUSR}
\end{figure*}

Consequently, the time-evolution of the ZF-$\mu$SR asymmetry spectra have been collected down to 0.03 K and the results are shown in Fig.~\ref{FIG:MUSR} (a) for some selected temperatures between 0.03 and 40 K. The absence of coherent spontaneous oscillations, expected for a log-range ordered ground state, and the lack of 1/3 polarization recovery, expected for a static disordered magnetic ground state, rule out any static magnetic ordering transition down to 0.03 K. All the ZF-$\mu$SR asymmetry curves from 40 to 0.03 K have been satisfactorily fitted using a single stretched exponential function,

\begin{equation}
    A(t) = A_{rel}\exp(-\lambda t)^{\beta}
\label{Musrfit}    
\end{equation}

$A_{rel}$ is the muon asymmetry amplitude which was obtained by fitting the 0.03 K data and then kept fixed at the same value of 0.2411 throughout the ZF data analysis. $\lambda$ and $\beta$ are, respectively, the relaxation rate corresponding to the electronic moment fluctuation and stretched exponent. As shown in Fig.~\ref{FIG:MUSR} (a), the asymmetry curves do not go through any noticeable change from the lowest measurement temperature of T = 0.03 K up to 1.5 K, and then upon further increase of temperature, the asymmetry curves encounter a gradual changeover by means of a systematic decrease in the muon spin relaxation. As demonstrated in Fig.~\ref{FIG:MUSR} (b), between 2 and 50 K, the muon spin relaxation rate, $\lambda$ gradually increases with decreasing temperature, suggesting a slowing down of Sm-spin-fluctuations through the development of magnetic correlations, as commonly reported in other QSL materials \cite{Abhisek_PRB_2024,Tushar_PRB_2017,Abhisek_PRB_2022,Kundu_PRL_2020}. Despite such an obvious slowing down of the spin dynamics, the SmTa$_7$O$_{19}$ material does not reveal any static magnetic ordering till down to the lowest measurement temperature of 0.03 K, as reflected from the missing of diverging relaxation rate until the lowest $T$ [see Figs~\ref{FIG:MUSR} (a) and (b)]. Notably, upon temperature lowering below $\sim$ 2 K, $\lambda$ levels-off ($\sim$ 0.3 $\mu$s$^{-1}$) and maintains a nearly temperature-independent plateau-like behavior between
2 K and 0.03 K [Fig.~\ref{FIG:MUSR} (b)], which is reminiscent of persistent strong quantum spin fluctuations~\cite{Abhisek_PRB_2024,Abhisek_PRB_2022,Kundu_PRL_2020,Kundu_PRL_2020_2,Li_PRL_2016}. This indicates that the correlated Sm$^{3+}$-moments in the lowest Kramers doublet state continue to fluctuate far below the magnetic interaction energy scale, $k_B\Theta_W$. The stretched exponent, $\beta$, gradually decreases from $\sim$ 1 at 50 K upon cooling down, reaching nearly a constant of $\beta \sim$ 0.6 between 2 and 0.03 K [see inset to Fig.~\ref{FIG:MUSR} (b)], reflecting emergence of an inhomogeneous magnetic environment at low-$T$ \cite{Nag_PRB_2018,Gauthier_2017}. One of the possibilities in our case is development of short-range magnetic correlations at this low-$T$ region, which could give rise to inhomogeneous magnetic environment and can affect $\beta$. However, it is to be noted that the value of $\beta$ ($\approx$ 0.6) at very low $T$ ($<$ 2 K) is higher than the $\beta$ = 1/3 of a canonical spin glass~\cite{Campbell_PRL_1994}, supporting further the absence of spin freezing in SmTa$_7$O$_{19}$. 

Before the development of plateau-like $T$-independent behavior of the relaxation rate, the observed temperature dependence of the ZF-$\mu$SR relaxation rate, $\lambda$, between 2 and 50 K [see Fig.~\ref{FIG:MUSR} (b)] can be described by the Orbach relaxation mechanism, associated with crystal-field fluctuations, as observed in other rare-earth based frustrated magnets including NdTa$_7$O$_{19}$~\cite{balents2010spin,arh2022ising}. We then fit the $T$-dependence of the ZF-muon spin relaxation rate using phenomenological model
\begin{equation}
    \frac{1}{\lambda} = \frac{1}{\lambda_0} + \frac{\eta}{\exp(\Delta_{ZF}/k_BT)}
\end{equation}
Here, the temperature-dependent term is the Orbach term while $\lambda_0$ $\sim$ 0.306(2) $\mu$s$^{-1}$ is the constant term and stands for the muon spin relaxation in the Kramers doublet ground-state. This behavior of the relaxation rate suggests that at temperatures below $\sim$2 K, effective spin-1/2 degrees of freedom emerge in the ground state, as the excited crystal-field states do not contribute to the muon spin relaxation at this low-$T$ region. On the other hand, the high-temperature ($>$ 1.5 K) relaxation rate is governed by a thermally activated behaviour with a gap, $\frac{\Delta_{ZF}}{k_B}$ $\approx$ 11.4 K, revealing that the lowest Kramers doublet is well separated from the excited Kramers doublets, which is also supported from our point charge model CEF calculation discussed in Sec. IIIC. It is worth noting that a very similar zero-field muon spin relaxation behavior with a low-$T$ plateau in the relaxation rate vs. $T$ plot was also found in several rare-earth based frustrated magnets \cite{arh2022ising,clark2019two,Khuntia_2023,Zorko_2008}.

Now in order to identify if the origin of the plateau-like temperature-independent zero-field muon spin relaxation between 2 K and 0.03 K is static (due to nuclear) or dynamic (due to electronic), we have performed muon decoupling experiments in several applied longitudinal fields (LFs) at $T$ = 0.1 and 4 K, and the results are shown in Fig.~\ref{FIG:MUSR} (c) and Appendix~\ref{Muon decoupling} for some selected LFs. The LF asymmetry curves were fit using the Eq.~\ref{Musrfit}, where $A_{rel}$ was fixed at the same value of 0.2411 as that for the ZF muon asymmetry fitting. As shown in the main panel of Fig.~\ref{FIG:MUSR} (d), the relaxation rate, $\lambda$, gradually decreases with increasing applied LF and finally attains a nearly unchanged finite value for $H_{LF} \geq$ 7.5 kOe. If the observed saturation of zero-field muon depolarization rate between 2 and 0.03 K arises from any static internal field of width $\Delta$$H$, the size of that static local field could be estimated as: $\Delta$$H$ = $\frac{\lambda_{ZF}}{\gamma_{\mu}}$ $\approx$ 3.5 Oe, where $\gamma_{\mu}$ = 2$\pi\times$135.5 MHz/Tesla is the muon's gyromagnetic ratio and $\lambda_{ZF}$ is the ZF-muon spin relaxation rate at the lowest measured 0.03 K. 
So, complete decoupling of the muon spins from the effect of static internal field could be anticipated in applied external longitudinal fields of (5-10)$\times$$\Delta$$H$~\cite{Abhisek_PRB_2024,Choi_PRL_2019,Tushar_PRB_2017,Kundu_PRL_2020,Kundu_PRL_2020_2,Srimanta_PRB_2021}. Here, the applied maximum longitudinal field is 15 kOe, which is about 4286 times larger than $\Delta$$H$ $\approx$ 3.5 Oe. Yet, astonishingly, the $\mu$SR asymmetry spectra exhibit noticeable relaxation and there is no sign of complete suppression of muon depolarization in the highest applied LF of 15 kOe [see Fig.~\ref{FIG:MUSR} (c) and $\lambda \sim$ 0.0012 $\mu$s$^{-1}$ at $H_{LF}$ = 15 kOe depicted in Fig.~\ref{FIG:MUSR} (d)], corroborating the existence of strongly quantum fluctuating Sm moments down to at least 0.1 K in SmTa$_7$O$_{19}$. 

The LF dependence of the muon spin relaxation rate has been analyzed using the Redfield formula in order to have a crude estimation of the magnitude and frequency of the fluctuating local internal field~\cite{Redfield}:
\begin{equation}
    \lambda_{LF}(H) = \frac{2\gamma_{\mu}^2<H_{loc}^2>\nu}{\nu^2 + \gamma_{\mu}^2H_{LF}^2}
\end{equation}
Here $\nu$ is the fluctuation rate (which is inversely proportional to the spin-spin correlation time,
$\tau$ = 1/$\nu$), $<H_{loc}>$ is the time average of the amplitude of the fluctuating local internal field, and $H_{LF}$ is the applied LF. As displayed in Fig.~\ref{FIG:MUSR} (d), our Redfield fit to the $\lambda$ vs. $H_{LF}$ curve gives rise to $<H_{loc}> \approx$ 17.6(3) Oe and $\nu \approx$ 18.9(9) MHz. Furthermore, as displayed in the inset of Fig.~\ref{FIG:MUSR} (d), the stretched exponent, $\beta$, initially exhibits small decrease, showing a weak minimum at $\beta \approx$ 0.42 around 75 Oe and then monotonically increases with LF and approaches gradually close to 1 with further increase of $H_{LF}$, implying field induced a more homogeneous distribution of the fluctuating local fields in this system. This is consistent with other candidate QSL materials ~\cite{arh2022ising}. 

Additionally, 4 K muon decoupling experiments were also performed and consequently, the collected $\mu$SR asymmetry spectra were analyzed using the same Eq.~\ref{Musrfit} and the results are displayed for some selected applied LFs in Appendix~\ref{Muon decoupling}. The observed LF dependence of the muon spin relaxation rate, $\lambda$, and the stretched exponent, $\beta$, obtained from the 4 K LF-data analysis, remain quite similar to the 0.1 K muon decoupling results. This, together with the weak $T$-dependence of the muon spin relaxation rates as a function of applied LF, further support rapid spin fluctuations and the absence of any spin ordering or freezing down to at least 0.03 K in SmTa$_7$O$_{19}$ despite sizable magnetic interactions~\cite{Kundu_PRL_2020_2}.   

\subsection{Specific heat}
\begin{figure*}[tbh!]
{\includegraphics[width=0.9\linewidth, trim={00 100 20 50},clip]{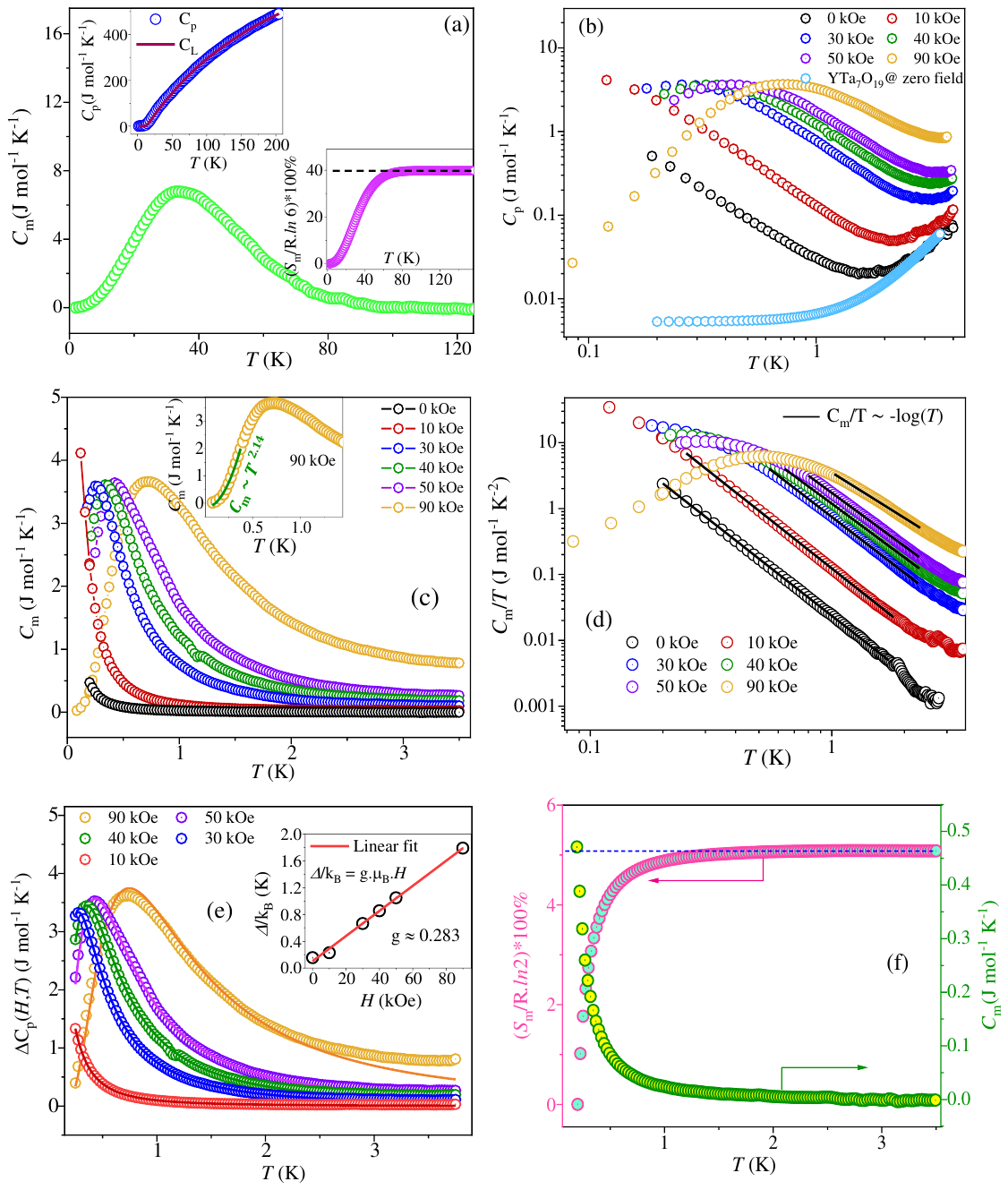}}
\caption{(Color online) (a) Temperature dependence of the zero-field magnetic specific heat between 2 and 125 K; Top left inset: Total measured specific heat ($C_p$) versus $T$ data along with the contribution from lattice part (solid red line); Bottom right inset: $T$-dependence of zero field magnetic entropy. (b) Total measured $C_p$ data of SmTa$_7$O$_{19}$ between 0.1 and 4 K in both zero and applied several magnetic fields, along with the zero-field $C_p$ vs. $T$ data of YTa$_7$O$_{19}$ in the $T$-range of 0.2-3.5 K. (c) Temperature dependence of the magnetic specific heat $C_m$ in the range of 0.1-4 K for different applied fields; Inset: 90 kOe $C_m$ versus $T$ plot along with the power-law fit (green solid line) for $T <$ 0.5 K. (d) $C_m/T$ versus $T$ curves on a log-log scale in the 0.1-4 K range along with the respective linear fittings. (e) Temperature variations of the [$C_p$($H$) - $C_p$($H$ = 0)] data at the respective applied $H$ along with the two-level Schottky anomaly fits (solid colored lines); Inset: Field dependence of the two-level Schottky energy gap (open black circles) and the linear fitting (solid red line). (f) $T$-dependence of the zero-field magnetic specific heat (right $y$-axis) and magnetic entropy (left $y$-axis) from 0.1 to 4 K.}
\label{FIG:SH}
\end{figure*}

An estimation of the magnetic entropy in any disordered magnetic material could also provide an alternative way to quantify the degree of magnetic frustration in that system. Therefore, to further examine the low-$T$ magnetic ground state and also to investigate the nature of low-energy spin excitations in SmTa$_7$O$_{19}$, we have measured the temperature dependence of zero-field specific heat ($C_p$) in the 2-300 K range and also between 0.1 and 4 K in both the zero and applied several magnetic fields. These results are shown in Figs.~\ref{FIG:SH} (a) and (b). As shown, 
there is no sharp $\lambda$-like anomaly down to 0.1 K at least, supporting the absence of thermodynamic phase transition into a long-range magnetically ordered phase and/or a structural phase transition in SmTa$_7$O$_{19}$. After subtracting the lattice part ($C_L$) from the total measured $C_p$ data using Debye-Einstein model (see details in Appendix~\ref{Lattice part estimation}), the resulting zero-field magnetic specific heat, $C_m$, is plotted 
in the main panel of Fig.~\ref{FIG:SH} (a) which shows a broad peak around 35 K, possibly indicating frustrated nature of magnetic interactions in SmTa$_7$O$_{19}$, as commonly discussed in the context of 
frustrated quantum magnets with competing exchange interactions ~\cite{balents2010spin,Abhisek_PRB_2022,Abhisek_PRB_2024,Abhisek_JPCM_2024,Cheng_PRL_2011}. Notably, the temperature region of this broad anomaly 
closely matches with the exchange energy scale, $k_B\Theta_W^{HT}$.

The release of zero field magnetic entropy, $S_m$ [see bottom right inset of Fig.~\ref{FIG:SH} (a)], was found to be only $\sim$ 40\% of the maximum $R$ln 6 ($\approx$ 14.9 J mol$^{-1}$K$^{-1}$) for a free Sm$^{3+}$ ion with $J_{eff}$ = 5/2 ground state. This possibly suggests the emergence of a CEF-split low-$T$ state, as also inferred from the DC susceptibility measurements. It is also evident 
that overall the magnetic entropy starts to decrease from $\sim$ 50 K with decreasing temperature, which is the similar temperature region below which the inverse susceptibility vs. temperature curve undergoes a slope change, revealing two distinct temperature regions for CW fitting, and also there is a cross over from the $J_{eff}$ = 5/2 ground state of a free Sm$^{3+}$ ion to the CEF-split low-energy state.

We now critically turn our focus to the low-$T$ (0.1-4 K) specific heat data. 
The magnetic specific heat, $C_m$, 
has been estimated after subtracting the $C_p$ of the nonmagnetic analog, YTa$_7$O$_{19}$ [see the zero field $C_p(T)$ data of YTa$_7$O$_{19}$ in Fig.~\ref{FIG:SH} (b)], and the resulting $C_m$-$T$ data of SmTa$_7$O$_{19}$ have been illustrated in Fig.~\ref{FIG:SH} (c) for different applied fields, showing no $\lambda$-like anomaly down to 0.1 K. As depicted in Fig.~\ref{FIG:SH} (d), the $C_m/T$ data in both zero and applied fields exhibit a negative log($T$) dependence with decreasing temperature, and then a broad peak-like feature developing in applied $H \geq$ 30 kOe [see Fig.~\ref{FIG:SH} (c)]. The negative log($T$) dependence of $C_m/T$ might suggest that the SmTa$_7$O$_{19}$ material could possibly lie in the quantum critical regime~\cite{Abhisek_PRB_2024,Abhisek_PRM_2024,Abhisek_JPCM_2024,Adroja_PRB_2022}. 

As our combined DC and AC magnetic susceptibilities, along with ZF, LF-$\mu$SR investigations, refute any static magnetic order down to 0.03 K, the low-$T$ anomaly in the $C_m$ vs. $T$ data is unlikely to arise from a magnetic transition. A more plausible explanation is that magnetic fluctuations in the Kramers doublet ground state of Sm$^{3+}$ dominate the low-$T$ specific heat in SmTa$_7$O$_{19}$, representing the onset of short-range correlations. 

In addition, the 90 kOe $C_m$ data yields a power-law dependence, $C_m$ $\propto \alpha T^n$, at very low temperature (below $\sim$0.5 K) with $n$ = 2.14(3) and a significantly large `$\alpha$' ($\sim$ 17.6 J mol$^{-1}$ K$^{-3.14}$) [see inset to Fig.~\ref{FIG:SH} (c)]. Such a $C_m(T)$ behavior in SmTa$_7$O$_{19}$ mimics the quadratic $T$-dependence $C_m$ and can be considered characteristic of gapless spin excitations from a QSL ground state, as is often observed in the spin-orbit-coupled heavier 4$d$ and 5$d$ transition metal-based oxides, serving as a fingerprint of a gapless QSL ground state \cite{Abhisek_PRB_2024,Abhisek_PRM_2024,Khuntia_PRB_2017,Trebst_2022}. It should also be noted that a perceptible power-law dependence of the $C_m(T)$ data is absent for the applied fields $H \leq$ 50 kOe. Therefore, specific heat measurements in those fields down to further lower temperatures are crucial to more critically probe the low-energy excitations.

As displayed in Fig.~\ref{FIG:SH} (c), with the increase of the applied magnetic field, the broad peak in $C_m(T)$ evolves into a two-level Schottky-type anomaly, which gradually moves towards higher temperatures with increasing $H$. This behavior is consistent with the field-induced splitting of the Sm-ground-state Kramers doublet. Furthermore, the temperature of the $C_m(T)$ peak linearly increases with increasing field, suggesting Zeeman splitting of the Sm-ground state doublet \cite{Adroja_PRB_2022}. A detailed discussion on ``two-level Schottky anomaly'' analysis has been discussed in Appendix~\ref{Lattice part estimation}, and the results are shown in Fig.~\ref{FIG:SH} (e). Based on the Schottky analysis, as depicted in the inset of Fig.~\ref{FIG:SH} (e), the Schottky energy gap, $\Delta/k_B$, maintains a linear relation with $H$, and the estimated $g$-value from the slope of this linear fit is found to be $\approx$ 0.283. It is worth highlighting at this point that while the energy gap between ground- and first excited state- Kramers doublet, estimated from the zero-field $\mu$SR characterization, is found to be $\Delta_{ZF}$/$k_B$ $\sim$ 11 K, the zero-field Schottky energy gap, $\Delta_0$/$k_B$ ($\sim$ 0.2 K), assessed from the field-dependent specific heat data analysis, turns out to be orders of magnitude smaller than $\Delta_{ZF}$/$k_B$. 
This clearly points to the fact that the low-$T$ Schottky anomaly in our SmTa$_7$O$_{19}$ is not due to the energy gap between ground- and first excited state Kramers doublets, rather it is likely to be attributed to the opening of the ground-state Kramers doublet of Sm$^{3+}$ with applied magnetic fields. On top of it, the small finite value of $\Delta_0$/$k_B$ at zero-field could be due to the splitting of the lowest Kramers doublet of Sm$^{3+}$ upon dynamic internal field~\cite{Adroja_PRB_2022} pertinent within the system down to the lowest 0.03 K. Another perspective could be posed here that at zero field, the increase of $C_m/T$ below $\sim$ 0.5 K (see Fig.~\ref{FIG:AP4} in Appendix~\ref{Lattice part estimation}) is due to the electronic spin excitations, which may exist in some quantum magnets where exchange interaction is weak, and dipole-dipole interaction is dominant to create an anisotropy \cite{Quilliam-thesis}. Similar to the reported quantum magnets Gd$_2$Sn$_2$O$_7$ \cite{Quilliam_PRL2007}, Yb(BaBO$_3$)$_3$ \cite{Jiang2022}, and Yb$_3$Ga$_5$O$_{12}$ \cite{Lhotel_PRB2021}, such an anisotropy could also induce a gap in the low-energy spin excitations of our SmTa$_7$O$_{19}$, which possibly gets reflected in the magnetic specific heat data analysis in terms of a non-zero Schottky energy gap, $\Delta_0$/$k_B$, at zero field.

\begin{figure}[tbh!]
{\includegraphics[width=0.9\linewidth, trim={150 150 190 70},clip]{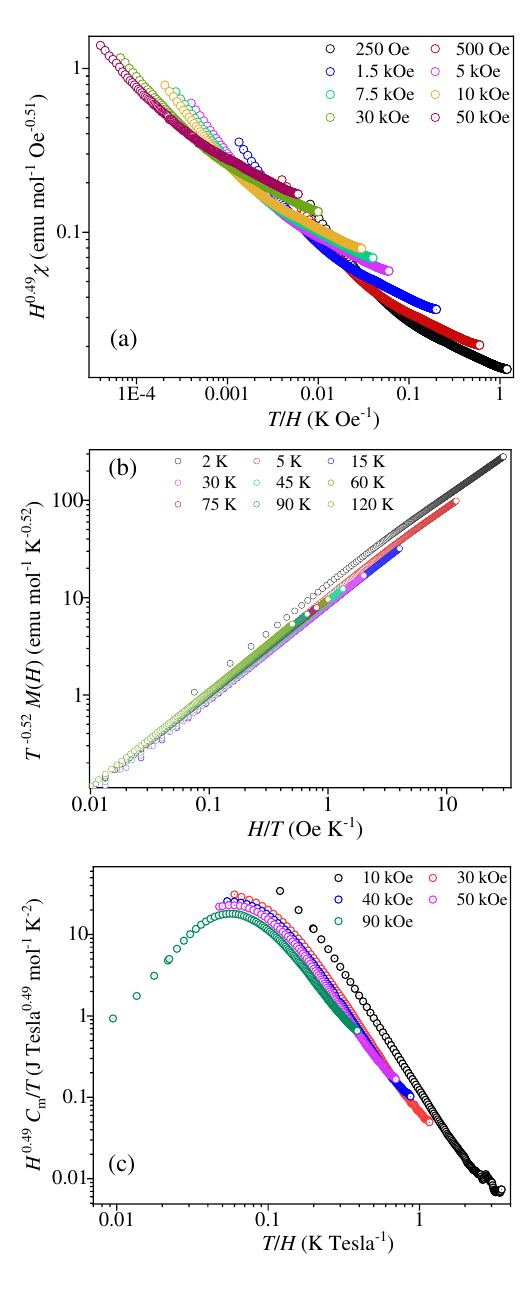}}
\caption{(Color online) (a) Scaling of $H^{0.49}$ $\chi_{DC}$($T$) with $T/H$ on a log-log scale. Log-log scaled plot of (b) $T^{-0.52}$ $M$($H$) versus $H/T$, and (c) $H^{0.49}$ $C_m/T$ versus $T/H$.}
\label{FIG:Scaling2}
\end{figure}

Finally, after eliminating the lattice and two-level Schottky contributions 
the $T$-dependence of the zero field magnetic specific heat, $C_m$, and the magnetic entropy, $S_m$, are shown in Fig.~\ref{FIG:SH} (f) between 0.1 and 4 K. The release of entropy is only about 5\% of the maximum $R$ln 2 $\approx$ 5.76 J mol$^{-1}$ K$^{-2}$ for a complete magnetic ordering of $J_{eff}$ = 1/2 Kramers doublet ground state of Sm$^{3+}$. This implies that most of the entropy is retained within the system till down to the lowest measured 0.1 K, supporting persistent strong quantum spin fluctuations in SmTa$_7$O$_{19}$. Notably, the magnetic entropy starts decreasing from around 1 K as the temperature decreases, which corresponds to the same temperature range as the $T$-independent plateau-like behavior in the ZF-$\mu$SR relaxation rate. All these evidences point towards the emergence of a gapless, dynamically fluctuating QSL-like ground state in SmTa$_7$O$_{19}$.

\subsection{Testing of Universal scaling behavior}
It is important to point out in this context that a random singlet state (RSS) has been regarded as a randomness-induced QSL, featuring common characteristics with true QSL. So, it is always a crucial issue to differentiate between subtle variations in states akin to QSL. Consequently, we have checked the validity of the universal scaling behavior of three distinct thermodynamic quantities, $\chi(T)$, $M(H)$, and $C_m/T$. These results are displayed in Figs.~\ref{FIG:Scaling2} (a) - (c), where we have plotted the $H^{\alpha_s} \chi(T)$ versus $T/H$ and $H^{\alpha_s} C_m/T$ versus $T/H$ curves at different applied fields, as well as the $T^{\alpha_m - 1} M(H)$ versus $H/T$ curves at different temperatures. As shown, unlike RSS, the ($H$, $T$) dependent $\chi_{DC}(T)$, $M(H)$, and $C_m/T$ data of our SmTa$_7$O$_{19}$ exhibit neither a universal scaling relation nor a data collapse into a single scaling curve with a similar value of the scaling exponent. This validates the presence of a dynamic QSL ground state in SmTa$_7$O$_{19}$, rather than an RSS as observed in Ba$_6$Y$_2$Rh$_2$Ti$_2$O$_{17-\delta}$~\cite{Suheon_PRR_2024}, Lu$_3$Sb$_3$Mn$_2$O$_{14}$~\cite{Suheon_PRB_2023}, and Li$_4$CuTeO$_6$~\cite{Khatua_2022} systems.       

\section{Conclusions}
We have investigated the magnetic ground state of a rare-earth triangular antiferromagnet using a comprehensive experimental study. Despite having antiferromagnetic exchange interaction between the Sm$^{3+}$ moments, our combined DC, AC magnetic susceptibility, specific heat, and $\mu$SR investigations evade any static magnetic ordering (neither a long-range magnetic ordering nor a spin-frozen ground state) down to 0.03 K. Further, our in-depth zero-field and longitudinal-field $\mu$SR data analysis establish a continually spin fluctuating dynamic magnetic ground state in SmTa$_7$O$_{19}$ down to 0.03 K. In addition, our scaling analysis of the $\chi(T)$, $M(H)$, and $C_m/T$ data negates a random singlet state, which is consistent with the disorder-free (in terms of site-disorder or bond-randomness) picture of SmTa$_7$O$_{19}$ as revealed by combined XRD and X-ray absorption studies. This supports the emergence of a low-temperature ($\sim <$ 2 K) quantum spin liquid phase in this compound. The $T$-divided magnetic specific heat, $C_m/T$, reveals an extensive negative logarithmic temperature dependence in the applied relatively lower fields, indicative of the emergence of quantum critical region in this material, while at the higher fields ($H \geq 40$ kOe), the $C_m$ follows a power-law dependence at very low temperatures ($\sim$ $<$ 0.5 K), referring to the existence of gapless spinon excitations in the QSL ground state of our SmTa$_7$O$_{19}$. 
The synergistic interplay between spin-orbit coupling and crystal electric field of the Sm$^{3+}$ ion, together with inherent geometrical frustration within the edge-shared equilateral Sm-triangular network, stimulate enhanced quantum fluctuations, preventing the system from magnetic phase transition and stabilizing a dynamic QSL ground state in SmTa$_7$O$_{19}$. Finally our work would certainly give way a fertile ground to test the novel quantum materials candidacy in this $RE$Ta$_7$O$_{19}$ family of triangular antiferromagnets.

\section{Acknowledgements}
The authors acknowledge the PSI muon facility for the $\mu$SR beam time on FLAME (Proposal ID 20231269) and Diamond light source for beam time on B18 beamline (SP34771-1). A. B. and D. T. A. thank Engineering and Physical Sciences Research Council, UK for funding (Grant No. EP/W00562X/1). D. T. A. would like to thank the Royal Society of London for International Exchange funding between the United Kingdom and Japan and Newton Advanced Fellowship funding between the United Kingdom and China. D.T. A. thanks the CAS for the PIFI fellowship. We thank Rafikul Ali Saha for the useful discussion on specific heat data analysis,  Matthias Gutmann on XRD analysis and Duc Manh Le on point charge model CEF calculation. DB and SE acknowledge funding from CSIR-HRDG through project file: 03WS (019)/2023–24/EMR-II/ASPIRE.

D.B. and A.B. contributed equally to this work.


\appendix

\section{\label{XANES}\texorpdfstring{Sm-\(L_3\) edge XANES comparison}{Sm-L3 edge XANES comparison}}
\begin{figure}[tbh!]
{\includegraphics[width=0.9\linewidth]{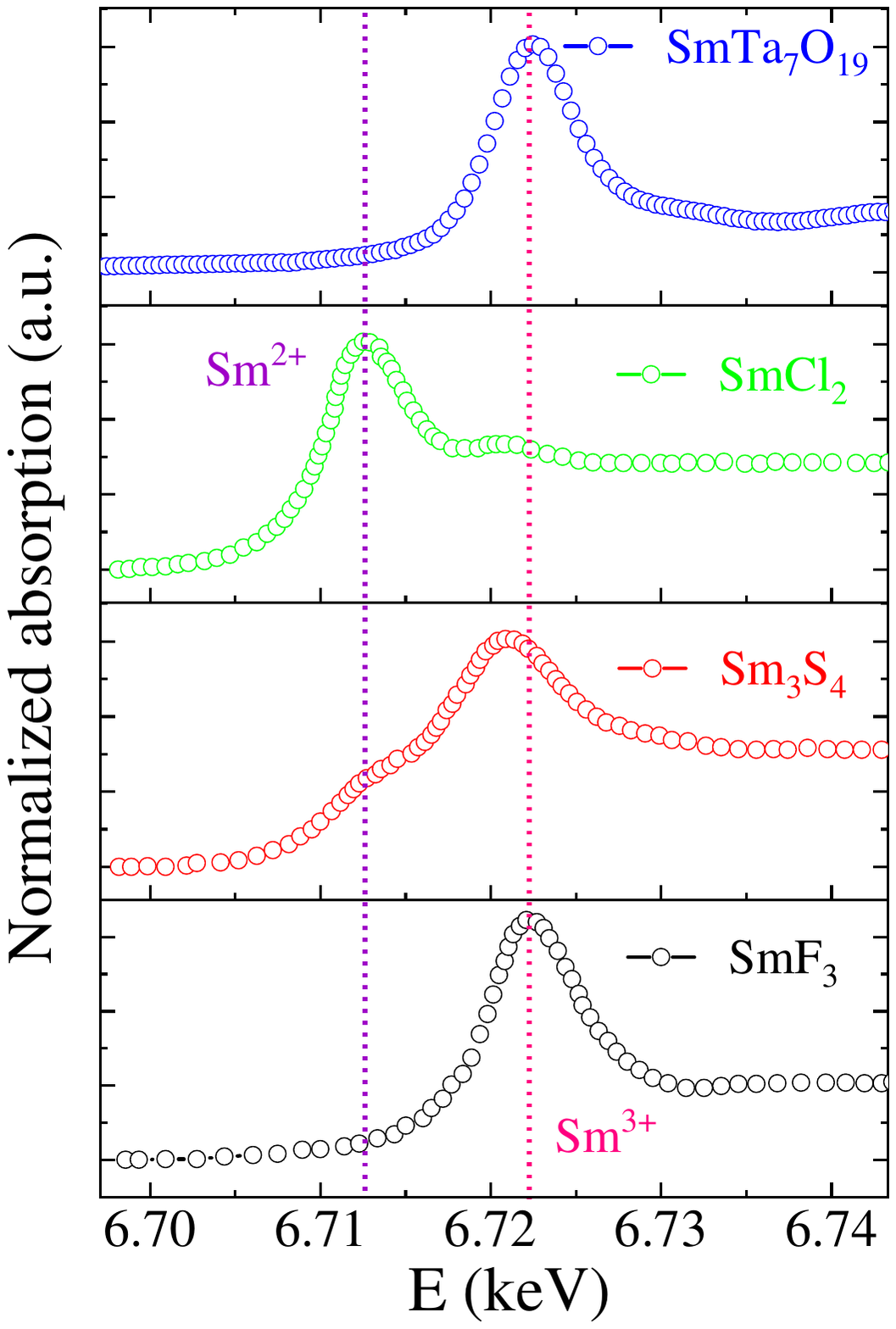}}
\caption{(Color online) Sm-$L_3$ edge XANES spectra of SmTa$_7$O$_{19}$, along with the reference divalent SmCl$_2$, trivalent SmF$_3$, and mixed di-/tri-valent Sm$_3$S$_4$. All three reference XANES spectra of the Sm-$L_3$ edge are adopted from ref.~\cite{Hu-JALCOM-1997} and digitized for making a comparative plot.}
\label{FIG:AP1}
\end{figure}

 Fig.~\ref{FIG:AP1} demonstrates a comparison of the Sm-$L_3$ edge XANES spectra between SmTa$_7$O$_{19}$ and the reference compounds containing distinctly different Sm-valence states. The vertical colored dotted lines correspond to the specific Sm-oxidation states for a guide to the eye.

\section{\label{Magnetic data}Additional magnetic results}
\begin{figure*}[tbh!]
{\includegraphics[width=1\linewidth]{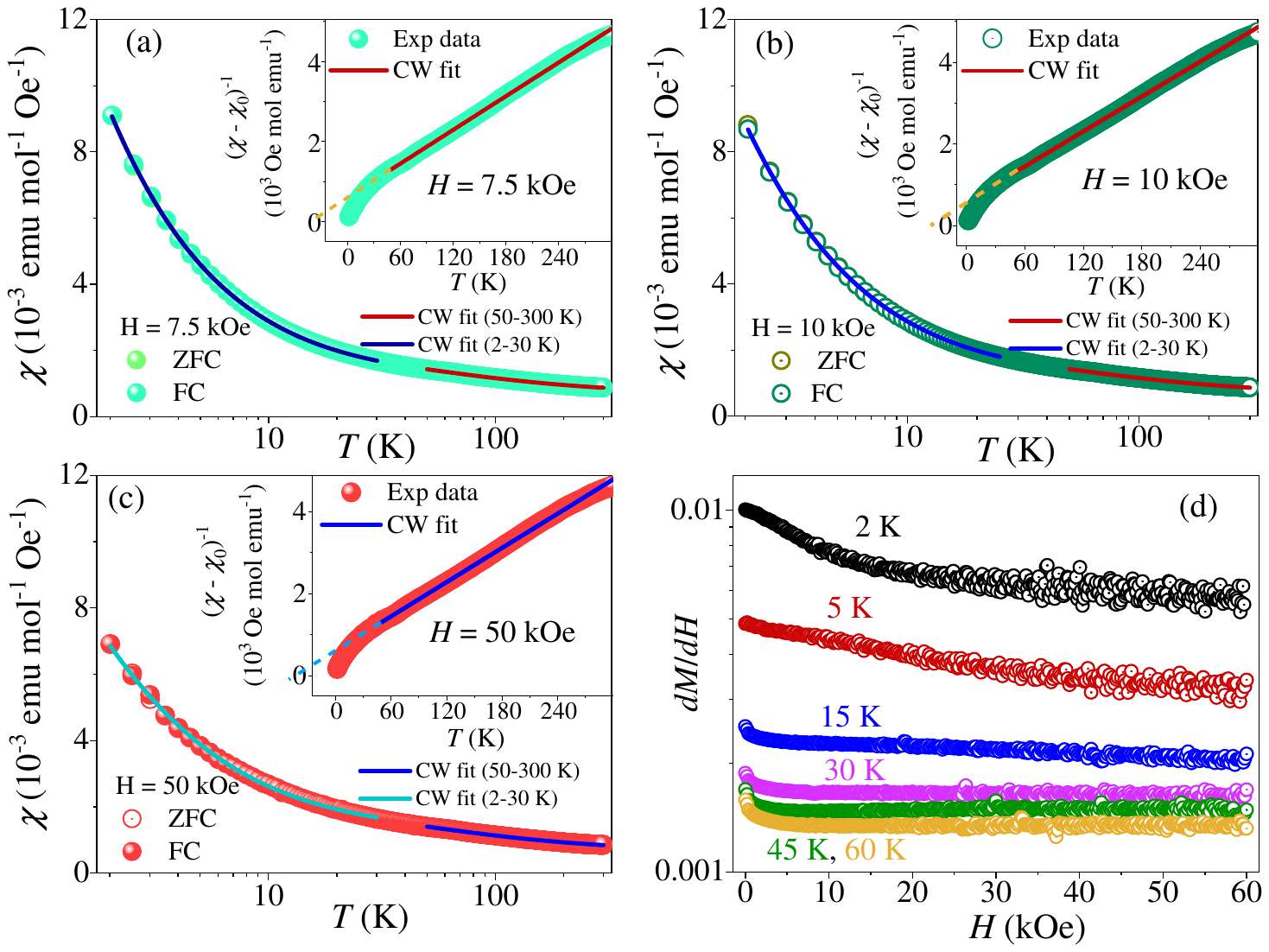}}
\caption{(Color online) (a) zero-field-cooled (ZFC) and field-cooled (FC) DC magnetic susceptibility curves as a function of temperature in 7.5 kOe (a), 10 kOe (b), and 50 kOe (c) applied fields along with the Curie-Weiss fits in both high-$T$ (50 - 300 K) and low-$T$ (2-30 K) regions; Insets of (a), (b), and (c) represent the respective inverse susceptibility versus temperature plots along with the linear CW fits. (d) Field derivative of DC magnetization as a function of applied magnetic field at different temperatures.}
\label{FIG:AP2}
\end{figure*}

 Fig.~\ref{FIG:AP2}(a-c) shows the temperature dependence of the ZFC and FC DC susceptibility data in different applied magnetic fields, along with the Curie-Weiss fits in both the high-$T$ (50-300 K) and low-$T$ (2-30 K) regions. Insets of Figs.~\ref{FIG:AP2}(a-c) display the linear Curie-Weiss fits to the inverse susceptibility versus $T$ between 50 and 300~K.

\begin{figure}
{\includegraphics[width=0.9\linewidth]{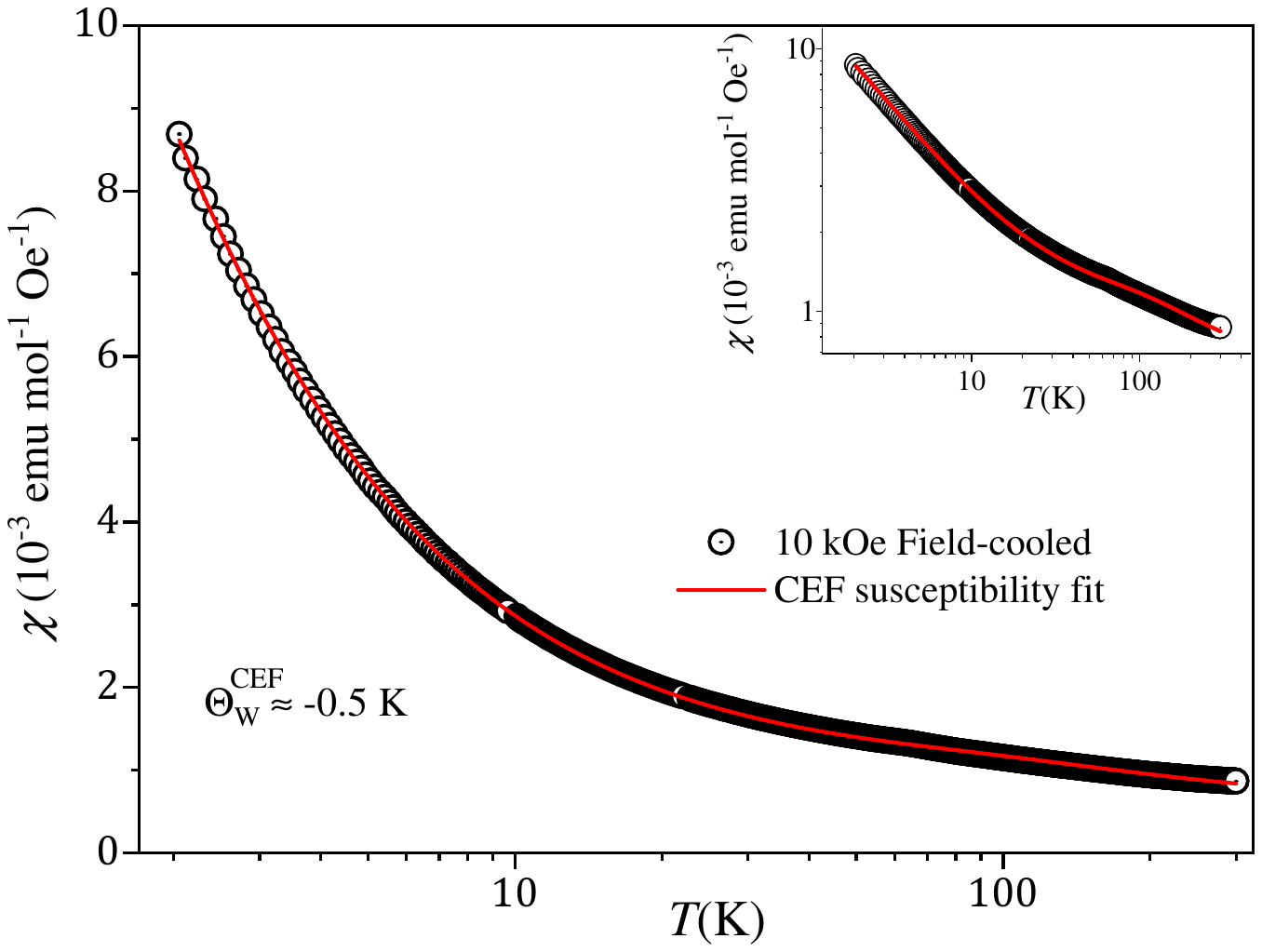}}
\caption{(Color online) 10 kOe field-cooled dc susceptibility data along with the CEF model fit (red solid line) \cite{CEFfit_Susc}; Inset: log-log scaled plot of the same for better illustration.}
\label{FIG:AP3}
\end{figure}

 Fig.~\ref{FIG:AP3} depicts the CEF model fit to the bulk dc susceptibility data over the full temperature range. Our fitting shows a very good agreement of the experimental data with the CEF model fit specifically at low temperatures (below $\sim$ 15 K).  

\begin{figure}
{\includegraphics[width=1\linewidth, trim={130 300 160 90},clip]{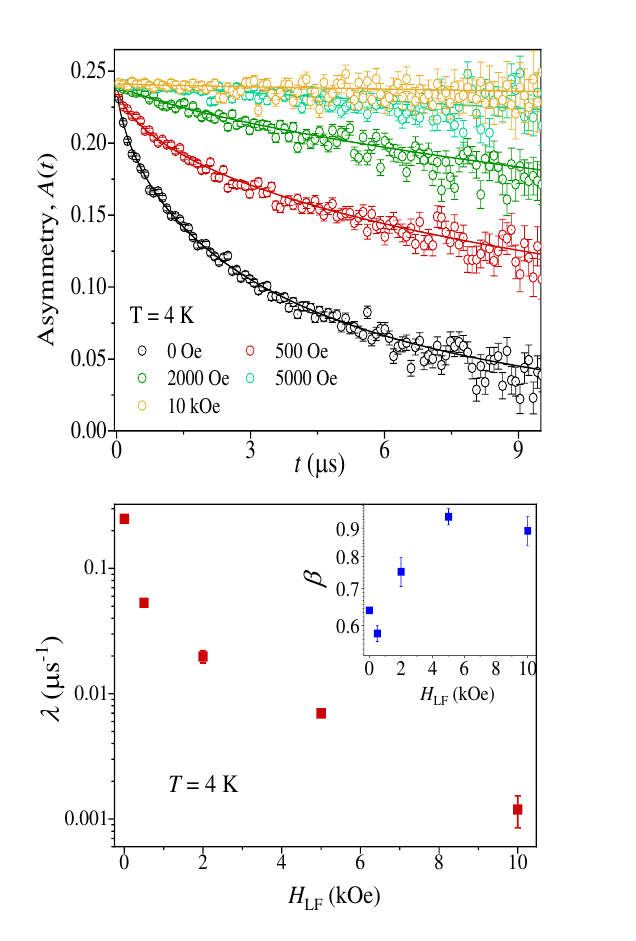}}
\caption{(Color online) (Top): Time evolution of the $\mu$SR asymmetry curves in several applied longitudinal fields at 4 K. (Bottom): Longitudinal-field dependence of the muon-spin relaxation rate and stretched exponent (Inset) at 4 K.}
\label{FIG:AP4}
\end{figure}
\begin{figure}[tbh!]
{\includegraphics[width=1\linewidth, trim={150 500 150 90},clip]{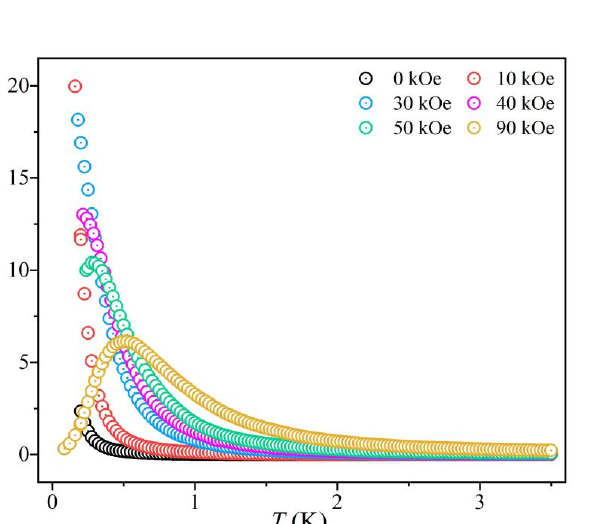}}
\caption{(Color online) $T$-divided magnetic specific heat versus temperature variations in both zero and applied several magnetic fields.}
\label{FIG:AP5}
\end{figure}

\section{\label{Muon decoupling}Additional longitudinal-field dependent muon asymmetry spectra}
Fig.~\ref{FIG:AP4} presents the results from the 4 K muon decoupling experiments, together with the collected data analysis. The corresponding fit parameters are also displayed as a function of applied longitudinal-field.

\section{\label{Lattice part estimation}Estimation of lattice contribution, and two-level Schottky anomaly analysis between 0.1 and 4 K}
 Fig.~\ref{FIG:AP5} presents the evolution of the $T$-divided magnetic specific heat data in applied magnetic fields, which clearly demonstrate Schottky anomaly behavior.

In order to obtain lattice contribution ($C_L$) to the total specific heat, we have fitted the zero-field $C_p$ data in the $T$-range of 100-300 K using a Debye-Einstein model with a combination of one Debye and three Einstein (1D + 3E) functions [see inset to Fig.~\ref{FIG:SH} (a)]. Our fitting result yields a Debye temperature $\Theta_D$ $\approx$ 100 K and three Einstein temperatures of $\Theta_{E_1}$ $\sim$ 298.6 K, $\Theta_{E_2}$ $\sim$ 294 K, and $\Theta_{E_3}$ $\sim$ 736.5 K. During fitting, $C_D$ and $C_{E_i}$ ($i$ = 1-3) were assigned as the weighting factors corresponding to the acoustic and optical modes, respectively, of atomic vibrations. Again, $C_D$ : $C_{E_1}$ : $C_{E_2}$ : $C_{E_3}$ = 0.037:0.24:0.34:0.383, resulting in the sum $C_D$ + $\sum_i$ $C_{E_i}$ $\approx$ 27 which is exactly the total number of atoms per formula unit in SmTa$_7$O$_{19}$, thus validating our lattice part fitting. This fit was then extrapolated down to the
lowest measured $T$ = 2 K and taken as the $C_L$, which
was then subtracted from the total $C_p$ to estimate the magnetic contribution of specific heat $C_m$.

The adopted strategy for analyzing the 0.1-4 K $C_p(T)$ data using a two-level Schottky anomaly model was as follows: first, we subtracted the zero-field $C_p$ data, i.e., $C_p$($H$ = 0), from those measured in the applied magnetic fields, i.e., $C_p$($H$ $\neq$ 0). Consequently, the $T$-dependence of this specific heat difference, $\Delta$$C_p(H,T)$, is illustrated in Fig.~\ref{FIG:SH} (e) and modeled using
\begin{equation}
    \Delta C_p(H,T) = [C_p(H \neq 0, T) - C_p(H = 0, T)]
\end{equation}
This was then fitted using a two-level Schottky anomaly model as
\begin{equation}
    \Delta C_p(H,T) = [C_{sch}(\Delta(H	\neq 0)) - C_{sch}(\Delta(H = 0))]
\end{equation}
$C_{Sch}(\Delta)$ and $C_{Sch}(\Delta_0)$ being the Schottky contributions to the specific heat. $\Delta(H)$ represents the Zeeman splitting in the applied magnetic field and $\Delta_0$ is the energy separation at $H$ = 0. Again, $C_{Sch}(\Delta)$ is defined as
\begin{equation}
    C_{Sch}(\Delta) = R(\frac{\Delta}{k_BT})^2\frac{\exp({\Delta/k_BT})}{[1 + \exp({\Delta/k_B})]^2}
\end{equation}
Correspondingly, the two-level Schottky anomaly fits were shown in Fig.~\ref{FIG:SH} (e).

\begin{table*}[]
\centering
\caption{A Few examples of rare earth based triangular lattice-based QSL materials. We have listed Curie-Weiss temperature, $\Theta_{W}$, frustration index, f, and the nearest neighbor R-R distance.}
\begin{tabular}{c c c c c c c}
\toprule
\textbf{Sample} & \textbf{Type of Lattice} & \textbf{\( \Theta_{CW} \) (K)} & \textbf{f} & \textbf{R-R distance(Å)} & \textbf{Ref} \\
\midrule
NaYbO\textsubscript{2} & Triangular & $-$ 10.3 & \(> 200 \) & 3.3507 & \cite{bordelon2019field} \\
YbMgGaO\textsubscript{4} & Triangular & $-$ 4 & \( > 80 \) & 3.4037&  \cite{paddison2017continuous}\\
TbInO\textsubscript{3} & Triangular (Honeycomb) & $-$ 1.17 & \( > 25 \) & 3.638 & \cite{clark2019two} \\
NaYbSe\textsubscript{2} & Triangular &  $-$ 3.5 & \( > 70 \) & 4.0568 & \cite{ranjith2019anisotropic, Gray2003} \\
CsNdSe\textsubscript{2} & Triangular & $-$ 0.66 & \( > 16 \) & 4.3468 & \cite{Xing2024} \\
Yb(BaBO\textsubscript{3})\textsubscript{3} & Triangular & $-$ 1.16& \( > 4 \)  & 5.4178 & \cite{Jiang2022} \\
PrMgAl\textsubscript{11}O\textsubscript{19} & Triangular & $-$ 8.1 & \( > 160 \)  &  5.584 & \cite{Ma2024}\\
PrZnAl\textsubscript{11}O\textsubscript{19} & Triangular & $-$ 8.9& \( > 180 \)  & 5.587 & \cite{bu2022gapless, Ashtar2019} \\
SmTa\textsubscript{7}O\textsubscript{19} & Triangular & $-$ 0.40 to $-$0.70 & \( > 13 \)  & 6.2101 & This compound \\
NdTa\textsubscript{7}O\textsubscript{19} & Triangular & $-$ 0.46 & \( > 11 \)  & 6.224 & \cite{arh2022ising} \\

\bottomrule
\end{tabular}
\label{tab:QSL materials}
\end{table*}

\newpage


\bibliography{apssamp}

\end{document}